\documentclass[%
 reprint,
 superscriptaddress,
 amsmath,amssymb,
 aps,
 pra,
]{revtex4-1}

\usepackage{graphicx}
\usepackage{dcolumn}
\usepackage{bm}
\usepackage{hyperref}
\usepackage{url}
\usepackage{comment}

\newcommand{\eref}[1]{Eq.~(\ref{#1})}

\newcommand{\fref}[1]{Fig.~\ref{#1}}
\newcommand{\Fref}[1]{Figure~\ref{#1}}

\begin{document}
\newcommand{\ManuscriptTitle}{
    A Collisional-Energy-Cascade Model \\
    for Nonthermal Velocity Distributions of Neutral Atoms in Plasmas    
}

\title{\ManuscriptTitle}


\author{Keisuke Fujii}
\email{fujiik@ornl.gov}
\affiliation{%
Oak Ridge National Laboratory, Oak Ridge, TN 37831-6169, United States of America
}

\date{\today}

\begin{abstract}
    Nonthermal velocity distributions with much greater tails than the Maxwellian have been observed for radical atoms in plasmas for a long time.
    Historically, such velocity distributions have been modeled by a two-temperature Maxwell distribution.
    In this paper, I propose a model based on collisional energy cascade, which has been studied in the field of granular materials. 
    In the collisional energy cascade, a particle ensemble undergoes energy input at the high-energy region, entropy production by elastic collisions among particles, and energy dissipation.
    For radical atoms, energy input may be caused by the Franck-Condon energy of molecular dissociation or charge-exchange collision with hot ions, and the input energy is eventually dissipated by collisions with the walls.
    I show that the steady-state velocity distribution in the collisional energy cascade is approximated by the generalized Mittag-Leffler distribution, which is a one-parameter extension of the Maxwell distribution.
    This parameter indicates the degree of the nonthermality and is related to the relative importance of energy dissipation over entropy production.
    This model is compared with a direct molecular dynamics simulation for a simplified gaseous system with energy input and dissipative wall collisions, as well as some experimentally observed velocity distributions of light radicals in plasmas.  
\end{abstract}



\maketitle

\section{Introduction}

Nonthermal velocity distributions have been observed for neutral atoms in plasmas for a long time~\cite{Vrhovac1991,Amorim2000,Samm1989,Hey1999,Shikama2004}.
Spectral profiles with much larger wings than a Maxwellian have been frequently observed.
While this is typically most apparent in hydrogen atomic emission lines, as their Doppler broadening is easily observed with a high-resolution spectrometer, similar nonthermal velocity distributions have been reported for other atoms~\cite{Sasaki2009-et}.
The origin of such a nonthermal velocity distribution has been attributed to generation processes of energetic atoms, such as Franck-Condon energy obtained through molecular dissociation, and charge-exchange collision with hot ions~\cite{Corrigan1965,Hey2004,Scarlett2017,McConkey2008,Starikovskiy2015, Sasaki2009-et}.
Many groups have empirically approximated these non-thermal energy distributions by a sum of two (or more) Maxwell distributions~\cite{Amorim2000, Hey1999, Sasaki2009-et}. 
However, the two-temperature model does not consider the relaxation of energetic atoms.
Furthermore, this model does not have a direct connection to a physical quantity, and thus it is difficult to extract knowledge from the observed nonthermal velocity distribution.
Some Monte-Carlo simulations also have reproduced the observed non-thermal velocity distribution~\cite{Sommerer1991,Starikovskiy2015,Ponomarev2017}, but it is not always applicable as all the physics quantities should be known beforehand for the system of interest.

In this paper, I propose to model such nonthermal velocity distributions of neutral atoms focusing more on the energy dissipation / relaxation processes than the energy input processes.
In particular, the application of the collisional-energy-cascade model is proposed, which has been studied in the field of granular gaseous.
This model has been originally proposed by Ben-Naim et al.~\cite{Ben-Naim2005-rd,Ben-Naim2005-uz,Kang2010-tk}, where they consider the three essential properties in the system;
1. a heat source in the high-energy limit, 
2. energy dissipation,
3. elastic collision among particles.
They point out that, under this condition, the steady-state velocity distribution has a power-law tail in the high-velocity region.
In an accompanying paper of this work~\cite{Fujii2022}, it is pointed out that this steady-state kinetic-energy distribution can be represented by the generalized Mittag-Leffler (GML) distribution, which is a one-parameter extension of the Maxwell distribution.
Although the GML distribution has no analytic representation except for a few special cases, its Laplace transform can be simply written by $\mathcal{L}_{f_{\operatorname{GML}}}(s) = \int_0^\infty f_{\operatorname{GML}}(E)\; e^{-sE} dE = [1 + 2D^{-1}(\langle E\rangle_\alpha s)^\alpha]^{-D/2\alpha}$. 
Here, $\langle E\rangle_\alpha > 0$  is the energy scale (see Ref.\cite{Fujii2022} for the relation to the fractional-calculus-extension of the mean energy) and $D$ is the spatial dimension of the system. 
$0 < \alpha \leq 1$ is the stability parameter related to the relative importance of the dissipation process, and
$\alpha = 1$ corresponds to the thermal system, where the GML distribution reduces to the Maxwell distribution. 
Thus, $\alpha$ can be seen as a dimensionless parameter representing the degree of nonthermality.

For radical atoms in plasmas, the heat source may be caused by the Franck-Condon energy of molecular dissociation or by charge-exchange collision with hot ions, while this input energy is eventually dissipated by the wall collisions.
Elastic collision among atoms may randomize the kinetic energy.
This similarity suggests the applicability of this collisional-energy-cascade model to the velocity distribution of radical atoms in plasmas, which is the purpose of this paper.
This paper is organized as follows.
In section \ref{sec:theory}, the kinetic theory of gaseous particles with energy dissipation and its connection to GML distribution~\cite{Fujii2022} is briefly summarized.
In section \ref{sec:md}, the direct molecular simulation for a simplified situation will be presented. In this simulation, the energy source, energy dissipation, and elastic collisions are taken into account.
The velocity distribution of particles is directly compared with the theoretical prediction.
In section \ref{sec:experiment}, several previous measurements for the velocity distribution of atoms in plasmas will be presented and compared with a GML distribution.

\section{Theory\label{sec:theory}}

In this section, a brief summary of the theory leading the GML distribution~\cite{Fujii2022} is shown. 
Also, a numerical computation of the GML distribution, as well as the velocity distribution corresponding to the GML energy distribution is described.

\subsection{Derivation of GML energy distribution for Maxwell gases}

Consider an isotropic and spatially uniform ensemble of particles undergoing elastic collisions (i.e., no energy dissipation at this point) in $D$-dimensional space. 
A Maxwell-type inter-particle interaction is assumed for now. Particle ensembles with other interactions will be discussed in subsection~\ref{subsec:nonmaxwell}. 
With a Maxwell interaction, the kinetic energies of two colliding particles, $E_1$ and $E_2$, can be thought of as random samples from the energy distribution $f(E)$.
For many collision systems, the post-collision energy $E_1'$ can be written by the following form~\cite{Fujii2022},
\begin{align}
  E_1' \leftarrow x E_1 + y E_2,
  \label{eq:recursive}
\end{align}
where $x, y \in[0,1]$ are random numbers following the probability distribution $p(x,y)$, which are determined by the collision geometry, such as the scattering angle and the relation between the relative and center-of-mass velocities. 
In steady state, $E_1'$ should also follow $f(E)$.

Several forms of $p(x, y)$ have been proposed. The simplest example of  valid $p(x,y)$ is the so-called \textit{diffuse} collision~\cite{Futcher1980-ey,Hendriks1982-cw}, where after the elastic collision the two energies will be completely randomized, i.e., no memory effect of pre-collision energies,
\begin{align}
p(x,y) = B\left( x \middle| \frac{D}{2}, \frac{D}{2}\right) \delta(x-y),
\label{eq:diffuse}
\end{align}
where $B(x|a, b) = x^{a-1} (1-x)^{b-1} / B(a, b)$ is beta distribution with beta function $B(a, b) = \int_0^1 x^{a-1} (1-x)^{b-1} dx$ and $\delta(t)$ is Dirac's delta function.
The $p$-$q$ model~\cite{Futcher1983-sf,Futcher1980-ey}, which takes the memory effect into account, as well as its linear superposition also gives a valid $p(x,y)$~\cite{Fujii2022}.

%
The steady-state solution of \eref{eq:recursive} can be written in the following form with the Laplace transform of the energy distribution $\mathcal{L}_f(s) \equiv \int_0^\infty f(E)e^{-sE}dE$, 
\begin{align}
    \mathcal{L}_f(s) =
    \int\, \mathcal{L}_f(xs) \mathcal{L}_f(ys)\, p(x, y) \,dx\,dy,
    \label{eq:laplace_nodissipation}
\end{align}
With any valid $p(x,y)$, the steady-state distribution converges to a Maxwell distribution $\mathcal{L}_f(s) = [1 + 2D^{-1}\langle E \rangle s]^{-D/2}$ according to Boltzmann's H-theorem. 
Here, $\langle E \rangle$ is the mean kinetic energy of the system.

Additionally consider a system with energy-dissipation.
It is assumed that, by this dissipation process, a particle loses its kinetic energy by the fraction of $1-e^{-\Delta}$ (with $\Delta \geq 0$).
The energy transfer to the surrounding walls can be considered as this dissipation process, but another process can be also considered.
A similar recursive relation with this dissipation can be constructed as follows
\begin{align}
    E_1' \leftarrow
    \begin{cases}
        e^{-\Delta} E_1, & \text{with probability}\ \xi\\
        x E_1 + y E_2, & \text{with probability}\ 1 - \xi
    \end{cases},
    \label{eq:recursive_dissipation}
\end{align}
where $\xi$ is the rate of this dissipation process relative to the elastic collision.
The Laplace representation of \eref{eq:recursive_dissipation} at the steady state is
\begin{align}
    \notag
    \mathcal{L}_f(s) &= 
    \xi \mathcal{L}_f(e^{-\Delta}s)\\
    &+(1 - \xi) \int\, \mathcal{L}_f(xs) \mathcal{L}_f(ys)\, p(x, y) \,dx\,dy.
    \label{eq:laplace_dissipation}
\end{align}
Here, it is implicitly assumed that constant energy injection exists in the high-energy limit so that the system will eventually arrive at a nontrivial steady state~\cite{Ben-Naim2005-rd}.

Consider the first two orders of $\mathcal{L}_f(s)$. 
From the normalization condition $\mathcal{L}_f(0) = 1$, it can be written that $\mathcal{L}_f(s)\approx 1 - (\langle E \rangle_\alpha s)^{\alpha}$ in the small-$|s|$ region, with $0 < \alpha \leq 1$.
Here, $\langle E \rangle_\alpha$ is the energy scale of the distribution, which is related to the fractional-calculus-extension of the mean energy~\cite{Fujii2022}.
Note that this corresponds to an assumption of $f(E)$ in the large-$E$ region, i.e., either $f(E) \approx E^{-\alpha-1} / (\langle E \rangle_\alpha)^{\alpha}\Gamma(1-\alpha)$ if $\alpha < 1$, or $f(E) \approx \exp(-E/\langle E \rangle_\alpha)/\langle E \rangle_\alpha$ if $\alpha=1$.
By substituting this into \eref{eq:laplace_dissipation}, we obtain
\begin{align}
    1 &= (1-\xi) \int (x^\alpha + y^\alpha) p(x, y)\, dx\, dy + \xi e^{-\alpha \Delta}.
    \label{eq:alpha}
\end{align}
Note that the symmetry of the elastic collision leads $p(x,y) = p(1-y, 1-x)$, which results in $\int (x+y) p(x, y) dx\, dy = 1$~\cite{Fujii2022}. 
This indicates that $\alpha = 1$ is the necessary and sufficient condition for the non-dissipative system, i.e., $\Delta = 0$ or $\xi = 0$.
With a finite energy dissipation, $\alpha < 1$.

At the large-$s$ limit, $\mathcal{L}_f(s)$ asymptotically behaves $\approx 2D^{-1}(\langle E \rangle_\alpha s)^{-D/2}$ at the steady state and with $\Delta \ll 1$~\cite{Fujii2022}.
By combining with its lowest order approximation 
$\mathcal{L}_f(s) \approx 1 - (\langle E \rangle_\alpha s)^\alpha$, 
we find that the generalized Mittag-Leffler (GML) distribution~\cite{haubold_mittag-leffler_2011,barabesi_new_2016,korolev_mixture_2020},
\begin{align} 
    \mathcal{L}_f(s) \approx \left[1+\frac{2}{D}\Bigl(\langle E \rangle_\alpha s\Bigr)^\alpha\right]^{-D/2\alpha},
    \label{eq:mittagleffler}
\end{align}
is the simplest approximation of the steady-state solution of \eref{eq:recursive_dissipation}.
The GML distribution naturally reduces to the Maxwell distribution at $\alpha\rightarrow 1$.

\begin{figure}
    \includegraphics[width=7.5cm]{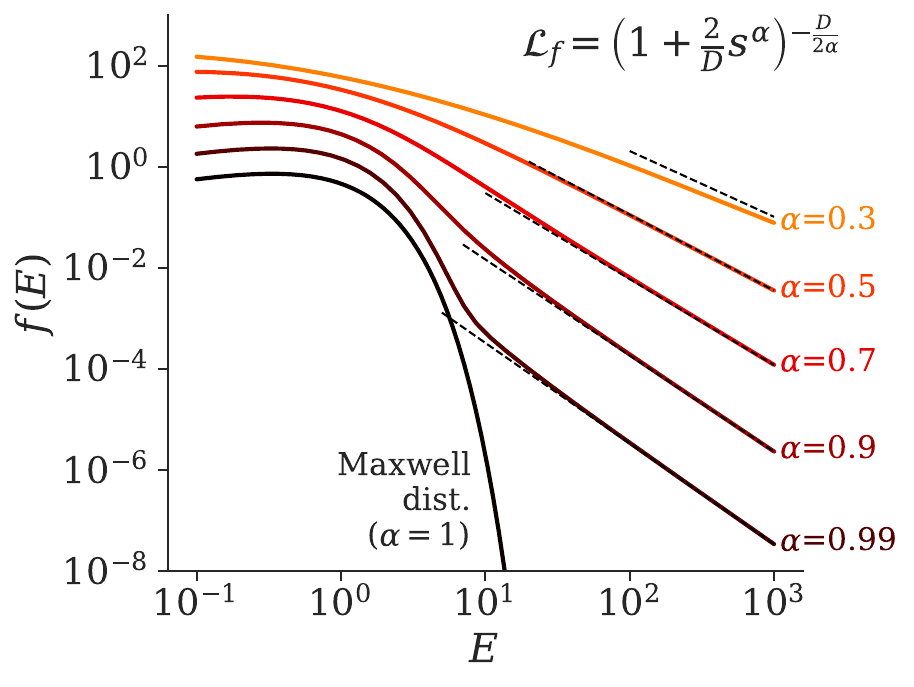}
    \caption{%
    The GML distribution with several values of $\alpha$ with $\langle E \rangle_\alpha = 1$. The power-law tail $E^{-\alpha-1}/\Gamma(1-\alpha)$ is shown by dotted lines.
    Appropriate offsets are introduced for clarity.
     }
    \label{fig:mittagleffler}
\end{figure}

\begin{figure}
    \includegraphics[width=7.5cm]{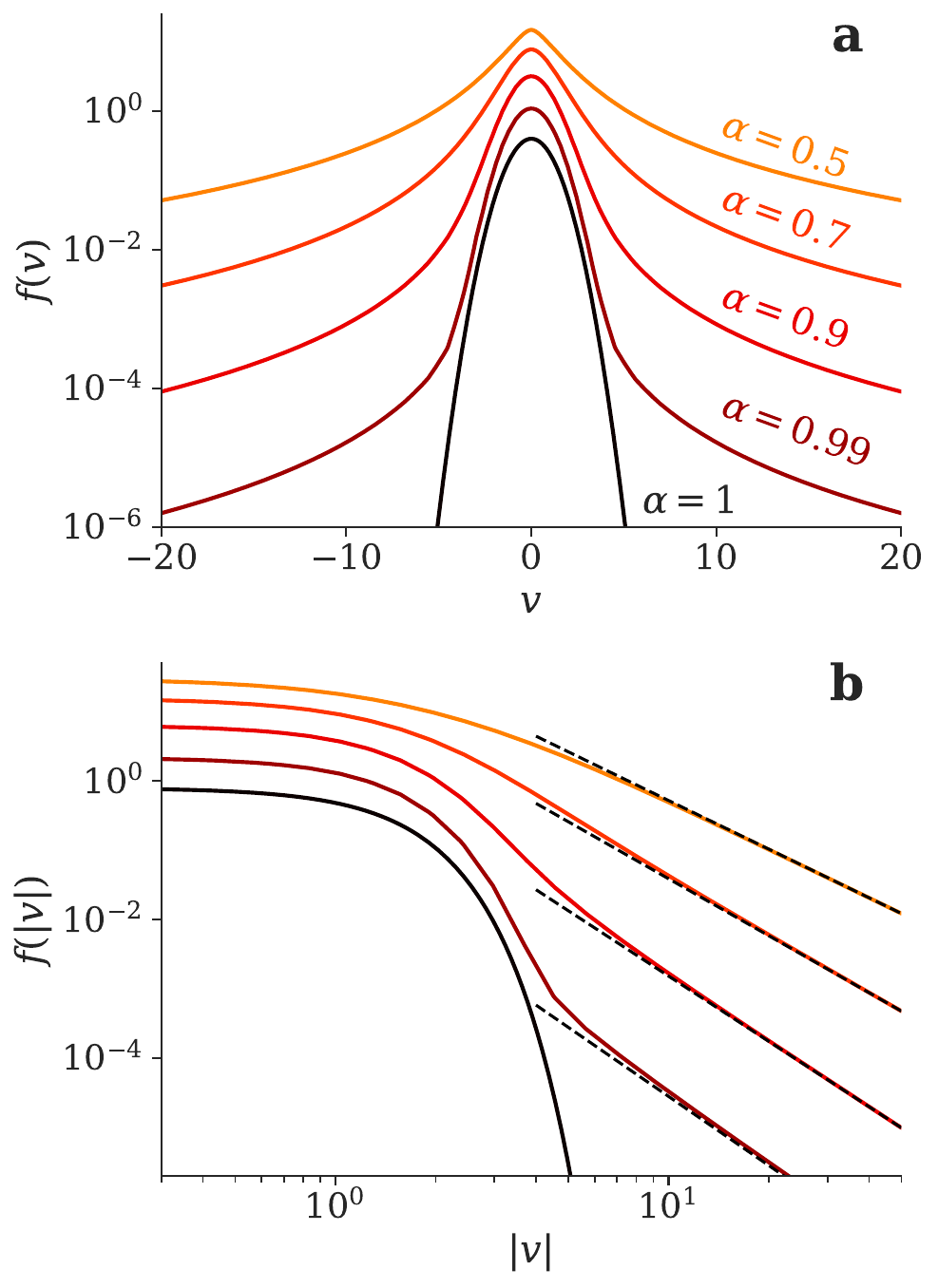}
    \caption{%
    Velocity distributions corresponding to GML energy distributions.
    $D=3$ and $\lambda=1/3$ are used.
    Distributions with several values of $\alpha$ are shown with vertical offsets for clarity.
    The distribution with $\alpha=1$ is a Gaussian.
    The distribution with $\alpha\lesssim 1$ is similar to a Gaussian in the central part but has much larger wings.
    With a smaller value of $\alpha$, the wing fraction is larger.    
    (a) is a semi-log plot and (b) is a log-log plot.
    The dotted lines in (b) shows the power-law distribution $\propto |v|^{-2\alpha - \lambda - 1}$.
    }
    \label{fig:full_gml_vdf}
\end{figure}

\subsection{Numerical evaluation of GML distribution}
Although the GML distribution has no analytical forms except for few special cases, some efficient numerical computation methods have been proposed~\cite{haubold_mittag-leffler_2011,barabesi_new_2016,korolev_mixture_2020}.
The GML distribution, $f_{\operatorname{GML}}(E|\alpha, D, \langle E \rangle_\alpha) = \mathcal{L}^{-1}\left[(1+2(\langle E \rangle_\alpha s)^\alpha / D)^{-D/2\alpha}\right]$ can be written as a mixture representation of the exponential function;
\begin{align}
    \notag
    &f_{\operatorname{GML}}(E|\alpha, D, \langle E \rangle_\alpha) = 
    \frac{1}{\pi \langle E \rangle_\alpha} 
    \left(\frac{D}{2}\right)^{\frac{1}{\alpha}}\\
    &\times\int_0^\infty \frac{
        \exp\left(-y\left(\frac{D}{2}\right)^{\frac{1}{\alpha}}\frac{E}{\langle E \rangle_\alpha}\right)
        \sin\left(\pi\frac{D}{2} F_\alpha(y)\right)
    }{
       \left(y^{2\alpha} + 2 y^\alpha \cos(\pi \alpha) + 1\right)^{D/4\alpha}
    }
    dy,
    \label{eq:gml_mixture}
\end{align}
where $F_\alpha(y)$ is defined as follows,
\begin{align}
    F_\alpha(y) = 1 - \frac{1}{\pi\alpha}\cot^{-1}\left(
        \cot(\pi\alpha) + \frac{y^\alpha}{\sin(\pi\alpha)}
    \right).
\end{align}
Observe that \eref{eq:gml_mixture} is the form of the weighted integration, $\int_0^\infty \exp(-cE) w(c) dc$, where $c$ is the scale and $w(c)$ is its weight (weight can be negative in this case).

\Fref{fig:mittagleffler} shows the GML distribution for several values of $\alpha$ with $D=3$ and $\langle E \rangle_\alpha=1$. 
Here, the function values of the GML distribution are computed by integrating \eref{eq:gml_mixture} numerically.
As expected from the small-$|s|$ dependence, it has a power-law tail, 
$E^{-\alpha-1}/\Gamma(1-\alpha)$.
The dotted lines in the figure are these power-law functions.
The GML distribution approaches to this power-law tail in the large-$E$ region.

The GML distribution approximates the distribution of the total kinetic energy.
With a spectroscopic measurement, only the velocity distribution along a particular axis is observed. 
In $D=3$ dimensional space, this velocity distribution can be evaluated by substituting $E = m(v_x^2 + v_y^2 + v_z^2)/2$ and integrate it over $v_x$ and $v_y$ by taking the statistical weight of free space into account, where $v_x$, $v_y$. and $v_z$ are the velocity components in the three dimensional space and $m$ is the mass of the particle.
Since only the term depending on $E$ in \eref{eq:gml_mixture} is $\exp(-cE)$ with $c = y\left(\frac{D}{2}\right)^{\frac{1}{\alpha}} / \langle E \rangle_\alpha$, the integration of this term is sufficient.

Consider the energy distribution $g(E | c) = c\exp(-cE)$ in 3-dimensional space.
As the statistical weight of the space is $\sqrt{2mE}$ in the energy domain, the distribution of $v_z$ is
\begin{align}
    \notag
    g(v_z | c)
    &= c\int_{-\infty}^\infty \exp(-cE) \frac{1}{\sqrt{2mE}}dv_x dv_y
    \\
    &= \frac{1}{2}\sqrt{\frac{c}{2m}}\;\Gamma\left(\frac{1}{2}, c\frac{v_z^2}{2m}\right),
    \label{eq:exponential_vdf}
\end{align}
where $\Gamma(s, x) = \int_x^\infty t^{s-1}e^{-t}dt$ is the lower incomplete gamma function.
By substituting \eref{eq:exponential_vdf} into \eref{eq:gml_mixture}, i.e., by replacing the term $\exp\left(-y\left(D/2\right)^{1/\alpha} E / \langle E \rangle_\alpha\right)$ in \eref{eq:gml_mixture} by $g\left(v_z \middle| y\left(D/2\right)^{1/\alpha} / \langle E \rangle_\alpha\right)$, we obtain the velocity distribution of particles with their total kinetic energy following the GML distribution.

\subsection{GML distributions for non-Maxwell gases\label{subsec:nonmaxwell}}

The above discussions can be approximately extended to particles having other inter-particle interactions.
For example, the collision rate of hard spheres is proportional to $E^{\lambda/2}$ with $\lambda=1$ while neutral atomic gases show Van-der-Waals interaction, where $\lambda=1/3$~\cite{Massey1934, Flannery2006}.
For such systems, we may consider the weighted distribution, $\hat{f}(E) = E^{\lambda/2} f(E) / Z$, with the normalization constant $Z$.
Based on an approximation $(E_1 + E_2)^{\lambda/2} \approx (E_1 E_2 / \langle E \rangle_\alpha)^{\lambda/2}$, which is valid if $|\lambda| \ll D$, this weighting approximately represents the energy dependence of the collision rate.
Although this weighting changes the statistical weight of the $D$-dimensional space from $\propto E^{D/2-1}$ to $\propto E^{(D + \lambda)/2 - 1}$, the Laplace transform of its weighted distribution at the steady state is approximated by the GML distribution, $\mathcal{L}_{\hat{f}}(s)=[1+(D + \lambda)(\langle E \rangle_\alpha s)^\alpha / 2]^{-(D+\lambda)/2\alpha}$.

For $\lambda \neq 0$ case, the energy distribution can be obtained by simply multiplying $E^{-\lambda/2}$ to \eref{eq:gml_mixture}.
\begin{align}
    \notag
    &f_{\operatorname{GML}}(E|\alpha, D, \langle E \rangle_\alpha, \lambda) \propto
    \frac{1}{\pi \langle E \rangle_\alpha} 
    \left(\frac{D+\lambda}{2}\right)^{\frac{1}{\alpha}}\\
    &\times\int_0^\infty \frac{
        E^{-\frac{\lambda}{2}} \exp\left(-y\left(\frac{D+\lambda}{2}\right)^{\frac{1}{\alpha}}\frac{E}{\langle E \rangle_\alpha}\right)
        \sin\left(\pi\frac{D+\lambda}{2} F_\alpha(y)\right)
    }{
       \left(y^{2\alpha} + 2 y^\alpha \cos(\pi \alpha) + 1\right)^{(D+\lambda)/4\alpha}
    }
    dy,
    \label{sup:eq:gml_mixture_x}
\end{align}
The corresponding velocity distribution can be obtained by replacing $g_\lambda(E | c) = cE^{-{\lambda/2}}\exp(-cE)$ by
\begin{align}
    g_\lambda(v_z | c)
    &= \frac{c^{\lambda/2}}{2}\sqrt{\frac{c}{2m}}\;\Gamma\left(\frac{1-\lambda}{2}, c\frac{v_z^2}{2m}\right).
    \label{sup:eq:exponential_vdf2_x}
\end{align}
To summarize, the velocity distribution corresponding to the energy distribution $E^{-\lambda/2} f_{\operatorname{GML}}(E)$ can be written by 
\begin{align}
    \notag
    f(v_z)
    \propto
    \int_0^\infty
    &g_\lambda\left(v_z \middle| y\left((D+\lambda)/2\right)^{1 / \alpha} \middle/ {\langle E \rangle_\alpha}\right)
    \\
    &\times
        \frac{\sin\left(
            \pi\frac{D+\lambda}{2} F_\alpha(y)
        \right)}{
           \left(y^{2\alpha} + 2 y^\alpha \cos(\pi \alpha) + 1\right)^{(D+\lambda)/4\alpha}
        }
        dy
    .
    \label{eq:GML_vdf}
\end{align}

\Fref{fig:full_gml_vdf} shows the velocity distribution corresponding to the GML energy distribution with $D=3$ and $\lambda=1/3$.
Distributions with several values of $\alpha$ are plotted.
With $\alpha=1$, the distribution is reduced to a Gaussian.
With $\alpha\lesssim 1$, the distribution has a similar profile to a Gaussian around the central region, though it has bigger wings.
With a smaller value of $\alpha$ (with larger energy dissipation), the wing intensity becomes bigger.

\Fref{fig:full_gml_vdf}~(b), the same profiles are shown in a log-log plot.
As expected from the energy distribution \fref{fig:mittagleffler}, the velocity distribution also has a power-law tail.
The dotted lines in the figure shows the power-law dependence $\propto |v|^{-2\alpha-\lambda-1}$.

\section{Direct Molecular Dynamics Simulation\label{sec:md}}

The previous discussion focuses on energy dissipation and relaxation (\eref{eq:recursive_dissipation}) and assumes that the energy source is at the high-energy limit.
However in realistic situations, there should be an energy cut-off at the energy scale of the heat source, and thus the power-law tail in \fref{fig:full_gml_vdf} only lasts up to this energy scale.
In order to see the validity of the previous discussion, as well as the effect of the energy cut-off, in this section a comparison is made to a simple molecular dynamics simulation.

\begin{figure}
    \includegraphics[width=7.5cm]{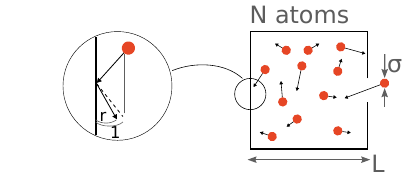}
    \caption{%
    A schematic illustration of our direct molecular dynamics simulation.
    $N$ atoms are confined by a cold box, where atoms will loose their kinetic energy by wall collisions.
    The box has a small opening and once an atom leaves the box through this opening, a high energy atom with temperature $T_0$ is injected.
    }
    \label{fig:md_illust}
\end{figure}

The system considered here is as follows (also illustrated in \fref{fig:md_illust});
\begin{enumerate}
    \item (energy randomization) $N$ atoms with mass $m$ are in a cubic box with one side of $L$.
    These atoms interact according to the inter-atomic potential $V(r) = (r/r_0)^{-6}$, where $r$ is the inter-atomic distance.
    \label{enum:randomization}
    \item (energy dissipation) The walls have infinitely large degrees-of-freedom and have much lower temperature than the atomic gas.
    A collision with the walls is approximated by an inelastic collision~\cite{Ito1985-px}, where the atomic velocity perpendicular to the wall changes $v_{\perp} \rightarrow -rv_{\perp}$ with the inelastic coefficient of the walls $r$. 
    $r=1$ indicates elastic collisions.
    \label{enum:dissipation}
    \item (heating) The box has an opening with the area of $a$. 
    If an atom goes out of the box through this opening, another atom having the temperature $T_0$ is injected into the box.
    \label{enum:heating}
\end{enumerate}
In steady state, this energy injection will be balanced with the energy dissipation by the wall collisions.     

\begin{figure}
    \includegraphics[width=7.5cm]{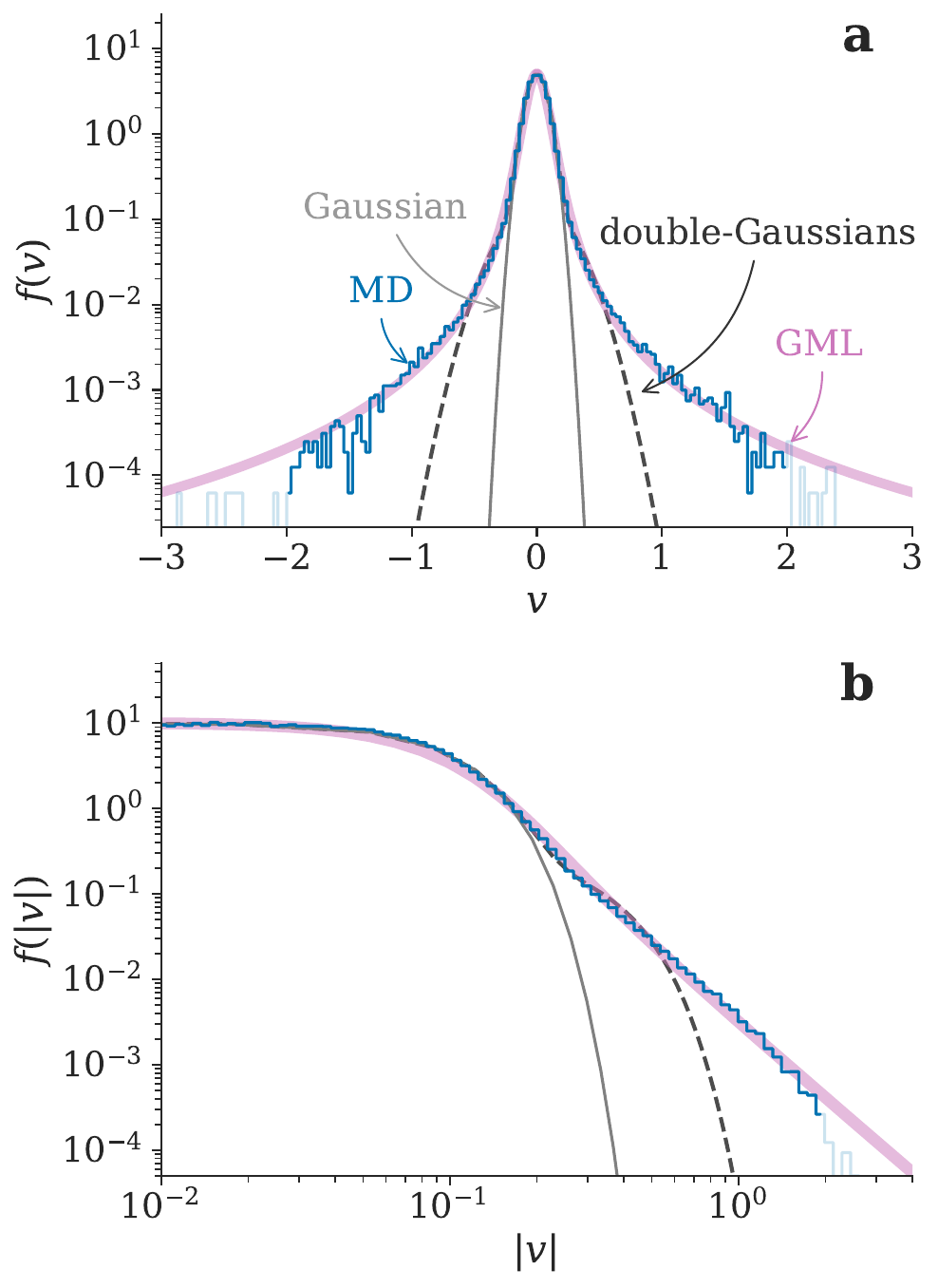}
    \caption{%
    (MD) The steady-state velocity distribution in the system shown in \fref{fig:md_illust} by the direct molecular dynamics simulation.
    $N=3\times 10^3$ and $r=0.99$ is used.
    (a) and (b) show the same distribution, but in a semi-log plot and log-log plot, respectively.
    The thin solid curve shows the best-fit by a Gaussian function.
    The simulated velocity distribution has a similar shape to a Gaussian around the low-velocity region, while in the high-velocity region it has a power-law dependence.
    The velocity of the heat source is $\approx 2$, forming a cut off of the power-law dependence.
    The bold curve shows the best fit by the GML velocity distribution~\eref{eq:GML_vdf}.
    The thin dashed curve shows the best fit by the double-Gauss function.
    }
    \label{fig:full_vdf}
\end{figure}

The steady-state velocity distribution of the atoms in this system is simulated with a molecular-dynamics simulator \texttt{lammps}~\cite{LAMMPS}. 
Used are $m=1$, $L=1$, $a=10^{-3}$, $r_0=10^{-3}$, $T_0=1$, and the time step of $2\times 10^{-5}$ for all the simulation results.
The values of $r$ and $N$ are scanned for the later comparison.
The simulation is continued until the system reaches the steady state.

\Fref{fig:full_vdf} shows the steady-state velocity distribution for $r=0.99$ and $N=3\times 10^3$.
The simulated velocity distribution has slowly decaying wings in high-velocity regions.
For a comparison, the best-fit Gaussian distribution is shown by a thin gray curve.
\Fref{fig:full_vdf}~(b) shows the same figure in a log-log plot.
It can be seen that the velocity distribution has a power-law tail in the high-velocity region up to $|v|\approx 2$, which roughly corresponds to the velocity scale of the heat source.

The bold curve in \fref{fig:full_vdf} shows the best-fit by GML velocity distribution.
Here $\lambda=1/3$ is used based on the Van-der-Waals inter-particle potential used in the simulation.
The GML distribution well captures the simulated velocity distribution, both the central and wing region, up to the velocity cut-off by the heat source.
Here, the number of adjustable parameters is two, which are the energy scale and $\alpha$. 
The optimum value of $\alpha$ at the best-fit is $0.903 \pm 0.002$ for this case.

As a reference, the best-fit by a double-Gauss function is shown by a dotted curve. 
Although the number of adjustable parameters for the double-Gauss fit (three, two energy scales and the intensity ratio of the two gaussians) is more than that in the GML fit, the double-Gaussian fails to represent the wings with the power-law decay.

\begin{figure}
    \includegraphics[width=7.5cm]{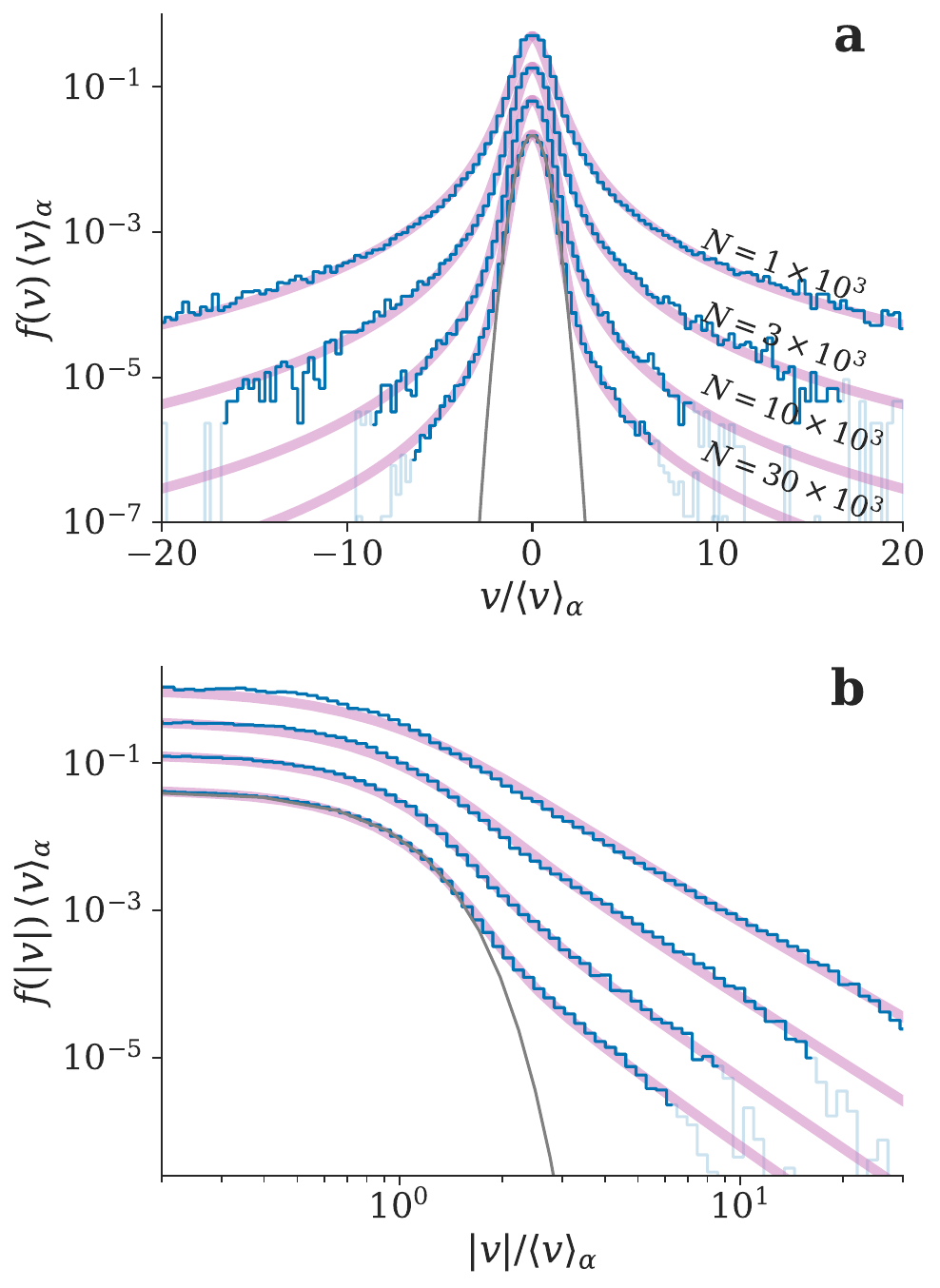}
    \caption{%
    Simulated steady-state velocity distributions with different values of $N$.
    The horizontal axis is normalized by the velocity scale $\langle v \rangle_\alpha$, while in the vertical direction an appropriate offset is introduced for the sake of clarity. 
    The intensity of the power-law tail is smaller to the larger-$N$ simulation, because of more significant entropy production by elastic collisions among particles.
    The bold curves show the best fit by the GML velocity distribution.
    The GML velocity distribution well captures the velocity distribution for all the conditions, particularly with large-$N$ systems.
    }
    \label{fig:full_vdf_several}
\end{figure}

\begin{figure}
    \includegraphics[width=7.5cm]{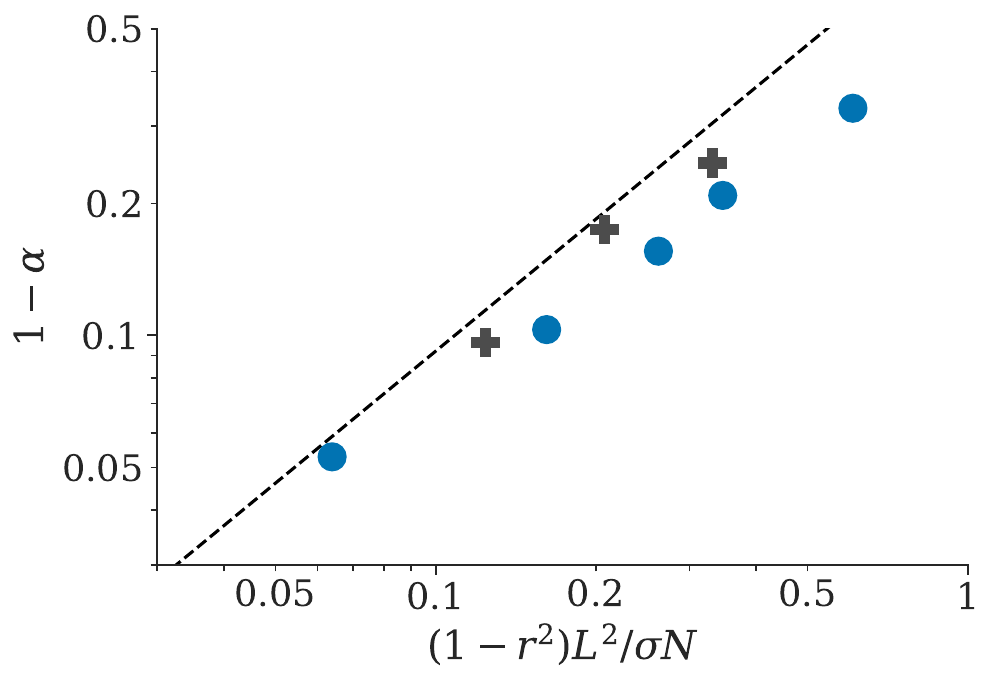}
    \caption{%
    $N$- and $r$-dependence of the best-fit $\alpha$.
    The vertical position for each simulation is computed from \eref{eq:alpha_md}.
    The circles and crosses indicate the simulation results with $r=0.99$ and $0.95$, respectively.
    The results are consistent with the theoretical prediction (dotted diagonal line), particularly with small dissipation condition.
    }
    \label{fig:full_md_alpha}
\end{figure}

Similar simulations are carried out with different values of $N$. 
The simulated velocity distributions are shown in \fref{fig:full_vdf_several}.
Note that the horizontal axis is the velocity normalized by $\langle v \rangle_\alpha \equiv \sqrt{2\langle E \rangle_\alpha / m}$ for clearer presentation of the shape difference.
Here, $\langle E \rangle_\alpha$ is obtained by the GML fit, which is described later.
With a larger number of particles in the box $N$, the profile becomes closer to a Gaussian distribution and the wing intensity becomes smaller.
This is consistent with the larger rate for the entropy production by elastic collisions.
The best-fit GML velocity distribution is shown by bold curves.  
The GML velocity distribution well captures the simulated velocity distribution. 
The quality of the fit is better at the smaller dissipation condition. 
This is consistent because the approximation \eref{eq:mittagleffler} is derived with $\Delta \ll 1$.
\Fref{fig:full_md_alpha} shows the $N$-dependence of the best-fit value of $\alpha$, as functions of $N$ and $r$.

The theoretical relation between $\alpha$ and other quantities may be derived from \eref{eq:alpha}.
The rate of the elastic collision is written by $\sigma \langle v\rangle_\alpha$, where as $\sigma$ the viscosity cross section at the energy scale $\langle E \rangle_\alpha$ is used, which is $\sigma \approx \pi r_0^2 (\langle E \rangle_\alpha)^{1/3}$~\cite{Baroody1961-vx}.
By substituting \eref{eq:diffuse} into \eref{eq:alpha} and $\xi \approx L^2 / \sigma N$, we obtain
\begin{align}
    1 - \alpha \approx \frac{(1-r^2)}{(D + \lambda)\left(\log 2 - \frac{1}{2 (D + \lambda)}\right)}
    \frac{L^2}{\sigma N}.
    \label{eq:alpha_md}
\end{align}
The dotted diagonal line in \fref{fig:full_md_alpha} shows \eref{eq:alpha_md}.
The simulated results with $r=0.99$ and 0.95 for several values of $N$ are close to this diagonal line.
This suggests the validity of the theory in the previous section.

\section{Experimental Observations\label{sec:experiment}}
In this section, the GML distribution is compared with experimentally observed velocity distributions of atoms in plasmas.
This GML velocity distribution is expected to be observed particularly for radical atoms in plasmas, because of the natural heating mechanism through molecular dissociation as well as the larger elastic-collision cross section through the Van-der-Waals interaction.

In order to observe the universality of the GML velocity distribution, it is compared with several experimental observations.
In subsection~\ref{subsec:amorim} and \ref{subsec:sasaki}, the velocity distributions of hydrogen atoms and oxygen atoms are analyzed, which have been observed with laser-based methods (Refs.~\cite{Amorim1996} and \cite{Sasaki2009-et}, respectively) and reported in the literature.
In their original publications, these distributions have been analyzed by a two-temperature Maxwellian.
It is shown here that the GML distribution also well represents the observations.
This suggests the universality of the collisional-energy-cascade in plasmas.
In subsection~\ref{subsec:nso_C}, unreported high-resolution emission spectra of neutral carbon and oxygen found in a public data archive~\cite{NSO} are studied.

\subsection{Hydrogen atom velocity distribution measured by laser-induced-fluorescence spectroscopy\label{subsec:amorim}}

\begin{figure}
    \includegraphics[width=7.5cm]{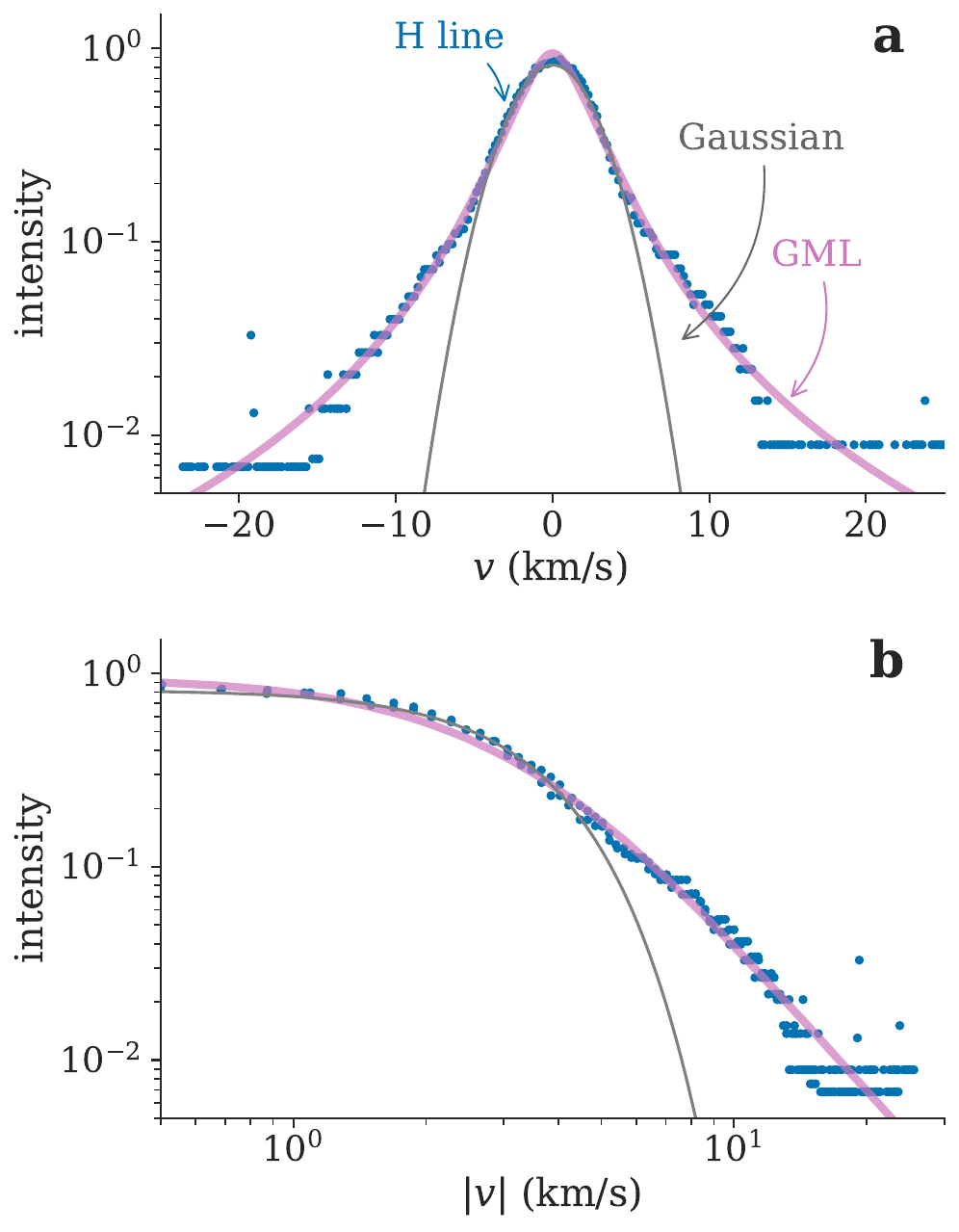}
    \caption{%
    The velocity distribution of hydrogen atoms in hydrogen--nitrogen mixture plasma observed by Amorim et al~\cite{Amorim1996-gj,Amorim2000}.
    The thin curve indicates the best-fit Gaussian, while the bold curve is the best-fit GML velocity distribution.
    The optimum value of $\alpha$ is 0.58.
    The high-velocity tail is well captured by the GML velocity distribution.
    The small discrepancy in the low-speed region may be due to the measurement resolution, which is $\approx$1.5 km/s.
    (b) is the same data for (a) but in a log-log plot.
    }
    \label{fig:full_amorim}
\end{figure}

Amorim et al. have directly measured the velocity distribution of ground-state hydrogen atoms in a microwave discharge tube with a hydrogen--nitrogen mixture~\cite{Amorim1996-gj,Amorim2000}.
They inject frequency-tripled 615-nm-laser (205 nm photon) into the plasma and the ground-state hydrogen atoms in the plasma are excited by two-photon absorption ($1s\; ^2\mathrm{S} \rightarrow 3d\; ^2\mathrm{D}$).
They observe the subsequent fluorescence ($3d\; ^2\mathrm{D} \rightarrow 2p\; ^2\mathrm{P}$, 656 nm transition).
From the fluorescence intensity as a function of the laser wavelength, the velocity distribution of the ground-state hydrogen atoms is obtained.
The resolution of the spectrum is determined by the line width of the 615-nm-laser, 0.08 cm$^{-1}$, which corresponds to 1.5 km/s in speed.
The data points are extracted from their electronic manuscript, where the data are embedded as an \texttt{xml} format, which is shown by markers in \fref{fig:full_amorim}.
The horizontal axis is the velocity along the sight line, converted from $v = c\Delta \lambda / \lambda_0$ where $c$ is the light speed, $\Delta \lambda$ is the wavelength displacement, and $\lambda_0$ is the line center.
The thin solid curve in \fref{fig:full_amorim} is the best-fit by a Gaussian function.
The central part of the velocity distribution is close to a Gaussian distribution, while it has much larger wings.
The wings profile is similar to a power-law dependence.

In their system, the energetic hydrogen atoms have been attributed to the photo-dissociation of ammonia~\cite{Amorim1996}.
These energetic atoms may distribute their energy to other particles by elastic collision. 
The injected energy should be eventually dissipated probably to the chamber walls, otherwise the atom temperature will increase up to the injection temperature in steady state. 
Therefore, their experimental condition is similar to that considered in this work: the existence of the energy injection at the high speed region, energy dissipation, and elastic collisions among particles. 

The bold curve in the figure shows the best-fit by the GML velocity distribution. 
Note that a small discrepancy in the low-speed region may be caused not only by the approximation introduced in \eref{eq:mittagleffler} but also the finite resolution of the measurement ($\approx 1.5$ km/s).
The GML velocity distribution well captures the observation, the Gaussian-like central part as well as the power-law decay in the tail, up to the energy cut-off around $|v| \approx 10^4$ m/s.
The optimum value of $\alpha$ obtained by the fit is 0.58.

From the optimum value of $\alpha$ and \fref{fig:full_md_alpha}, we obtain $(1 - r^2) L^2 / \sigma N \approx 1$.
From the chamber diameter $L\approx 0.016$ m, the elastic collision cross section among hydrogen atoms $\sigma \approx 3\times 10^{-19}\,\mathrm{m^{2}}$~\cite{international1999iaea}, and the expected atom density in the plasma $NL^{-3} \approx 10^{19} \,\mathrm{m^{-3}}$~\cite{Amorim1996}, the value of $r$ is estimated as $\approx 0.98$.
This is in a reasonable order for the inelasticity of a solid surface~\cite{Ito1985-px}, 
although in this rough estimate we ignored the collision with other species, such as collisions with hydrogen molecules.

\subsection{Oxygen atom velocity distribution measured by laser absorption spectroscopy~\label{subsec:sasaki}}
\begin{figure}
    \includegraphics[width=7.5cm]{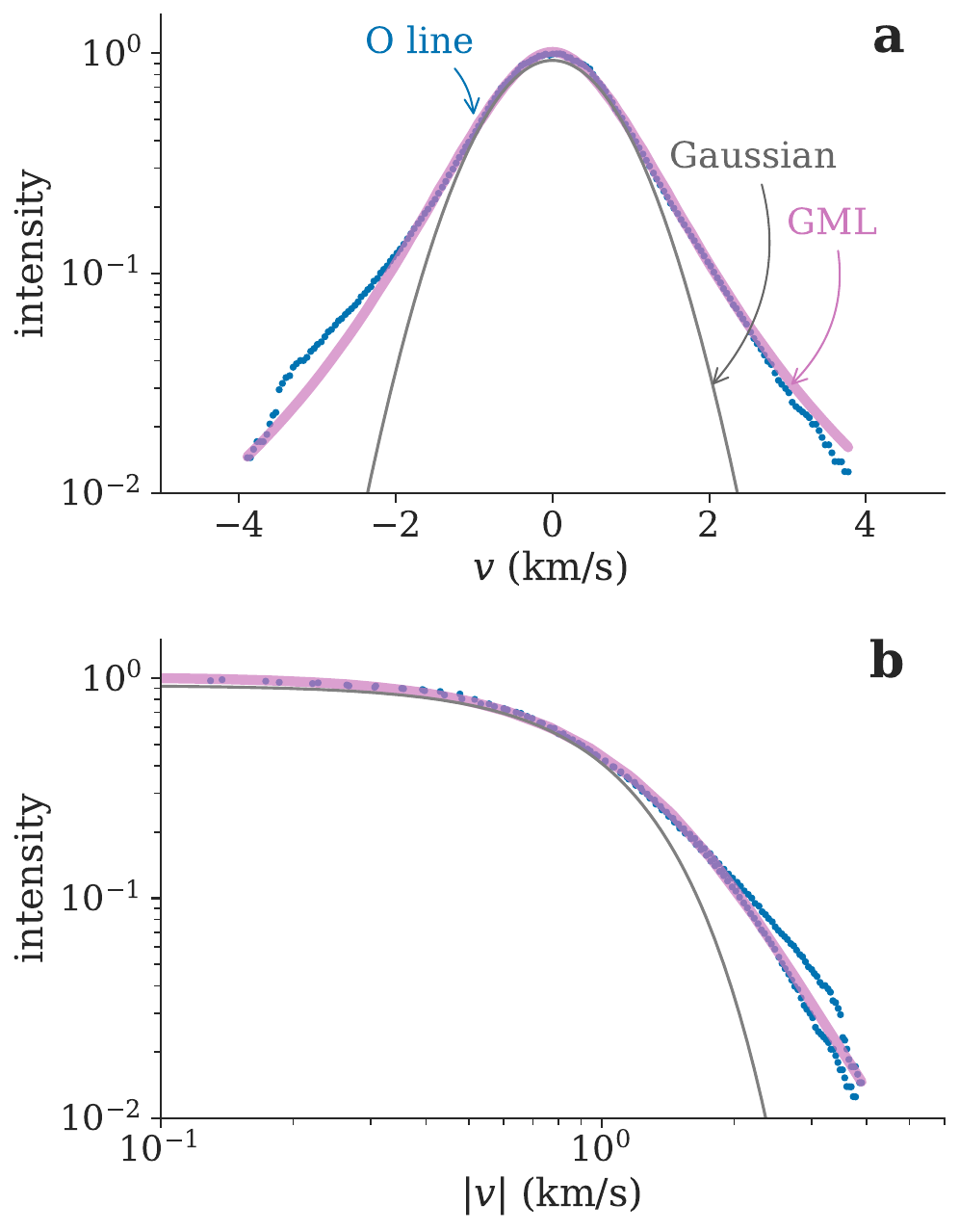}
    \caption{%
    The velocity distribution of oxygen atoms in helicon-wave-heated plasma of oxygen, observed with absorption spectroscopy based on a semiconductor-diode laser, reported by Sasaki et al~\cite{Sasaki2009-et}.
    The thin curve indicates the best-fit Gaussian, while the bold curve is the best-fit GML velocity distribution.
    The optimum value of $\alpha$ is 0.73.
    The central part and the high-velocity tails are well captured by the GML velocity distribution.
    The discrepancy in the far wings is attributed to the error in the zero level in the absorption spectrum.
    }
    \label{fig:full_sasaki}
\end{figure}

Sasaki et al. have measured the velocity distribution of metastable-state oxygen atoms in a helicon-wave heated plasmas of oxygen~\cite{Sasaki2009-et}.
The back-filled gas pressure and the helicon-wave power in their experiment shown here are 30 mW and 1.5 kW, respectively.
They inject a semiconductor-diode laser at 777.5 nm into a plasma and observe the absorption by the metastable oxygen transition $2s^2 2p^3(^4\mathrm{S^o})3s\; ^5\mathrm{S^o}_2 \rightarrow 2s^2 2p^3(^4\mathrm{S^o})3p\; ^5\mathrm{P}_1$.
By scanning the laser frequency, the absorption spectral profile (in this case the Doppler broadening is dominant) is obtained.

\Fref{fig:full_sasaki} shows the velocity distribution of oxygen atoms, which is extracted from their electronic manuscript. 
The horizontal axis is the velocity along the sight line, converted from $v = c\Delta \lambda / \lambda_0$ where $\lambda_0 = 777.5$ nm is the line center.
As the line width of the diode laser is much smaller than the Doppler width of oxygen in room temperature, the effect of the instrumental profile is virtually negligible.
On the other hand, an accurate intensity measurement at the far wings is more difficult to obtain in absorption spectroscopy.

The thin solid curve in \fref{fig:full_amorim} is the best-fit by a Gaussian function.
The central part of the velocity distribution is close to a Gaussian distribution, while it has much larger wings.
In their system, it has been suggested that the energetic oxygen atoms have been attributed to electron-impact dissociation of oxygen molecules~\cite{Sasaki2009-et}, similarly to the hydrogen atoms discused in the pevious subsection.
Originally, this nonthermal velocity distribution has been analyzed by a two-temperature Maxwellian.

The bold curve in the figure shows the best-fit by the GML velocity distribution. 
The GML velocity distribution well captures the observation.
The optimum value of $\alpha$ obtained by the fit is $0.73 \pm 0.01$.
From the optimum value of $\alpha$ and \fref{fig:full_md_alpha}, we obtain $(1 - r^2) L^2 / \sigma N \approx 0.3$.
Their chamber diameter is $L\approx 0.016$ m, and the oxygen density in the plasma is estimated from the back-fill pressure 30 mTorr and the dissociation ratio, and energy scale $\approx 1000$ K to be $NL^{-3} \approx 3\times 10^{20} \,\mathrm{m^{-3}}$.
From the elastic collision cross section among oxygen atoms $\sigma \approx 2\times 10^{-19}\,\mathrm{m^{2}}$~\cite{Mankodi2020-ve}, the value of $r$ is estimated as $\approx 0.84$.
This is again in a reasonable order for the inelasticity of a solid surface~\cite{Ito1985-px}.

\subsection{Carbon and oxygen atom velocity distributions measured by emission spectroscopy~\label{subsec:nso_C}}

\begin{figure*}
    \includegraphics[width=15cm]{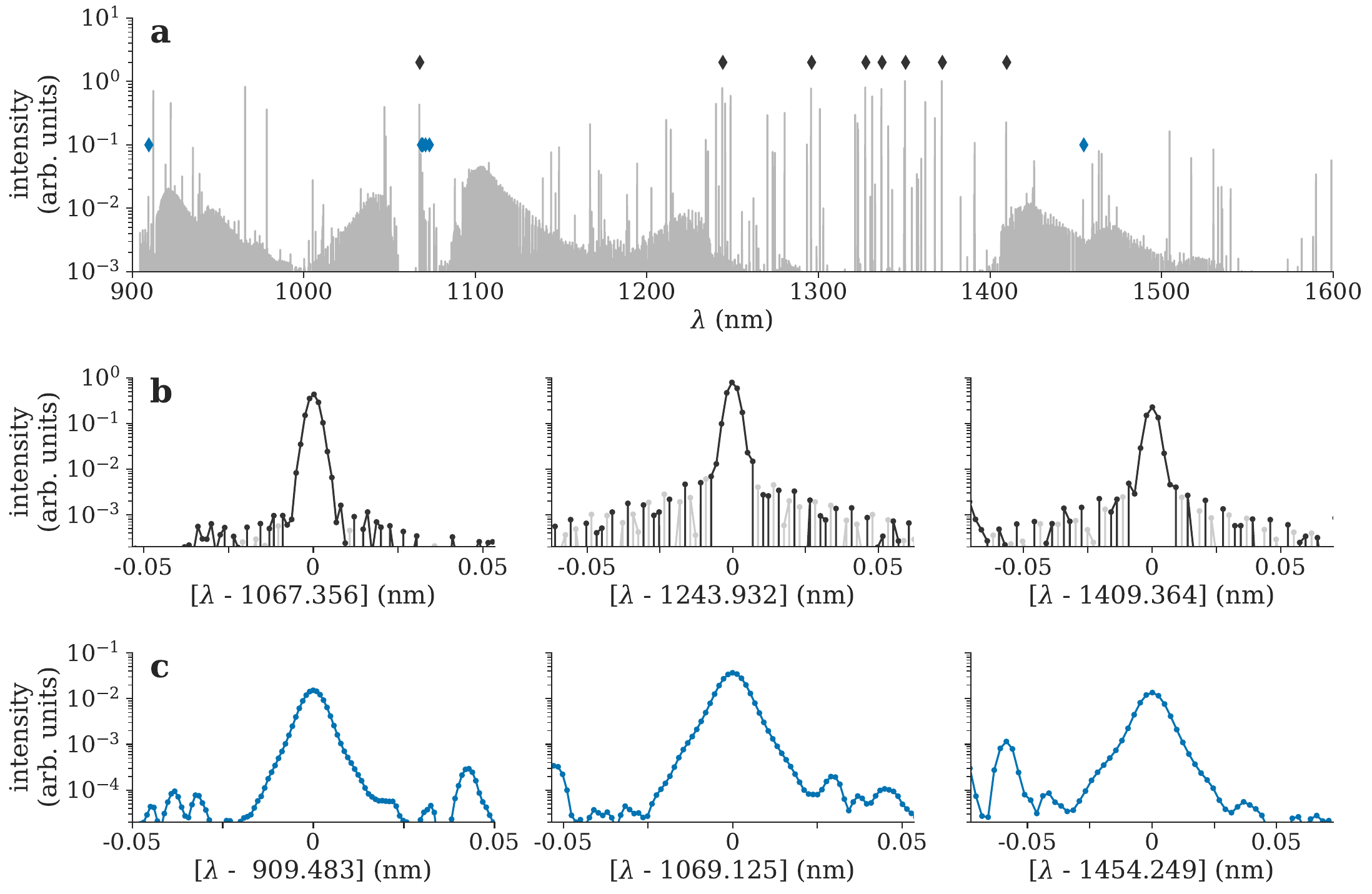}
    \caption{%
    (a) Spectrum observed in NSO~\cite{NSO}. 
    Grey and blue markers show the wavelengths of the argon and carbon lines, respectively, shown in \Fref{fig:full_nso_C}. 
    (b) Expanded spectra of three argon lines.
    The light gray markers show the negative points (multiplied by -1), representing the oscillation of the instrumental side robes.
    (c) Expanded spectra of three carbon lines.
    }
    \label{fig:full_nso_Call}
\end{figure*}

As a third example, consider a high-resolution spectrum measured at the National Solar Observatory, the electronic data of which are obtained from its historical data archive (file name \texttt{770612R0.010} \cite{NSO}).
This spectrum was originally measured to study the cyanide spectra in 1977 from a microwave discharge with a mixture of carbon, nitrogen, and argon, with a high-resolution Fourier transform spectrometer with a 1-m optical path difference in the wavelength range $\lambda=$ 830--2710 nm. 
The wavelength resolution is 3.5 pm (as the full width half maximum) for this measurement.
\Fref{fig:full_nso_Call}~(a) shows the original data and \fref{fig:full_nso_Call}~(b) and (c) show expanded views for 
argon lines at $\lambda =$ 
1067.36 nm ($3s^23p^5(\mathrm{^2P^o_{3/2}})4p\; \mathrm{^2[1/2]_1} \leftarrow 3s^23p^5(\mathrm{^2P^o_{3/2}})5s \; \mathrm{^2[3/2]^o_2} 	$), 
1243.93 nm ($3s^23p^5(\mathrm{^2P^o_{3/2}})4p\; \mathrm{^2[1/2]_1} \leftarrow 3s^23p^5(\mathrm{^2P^o_{3/2}})3d \; \mathrm{^2[3/2]^o_2} $), 
and 
1409.36 nm ($3s^23p^5(\mathrm{^2P^p_{3/2}})4p \; \mathrm{^2[1/2]_0} \leftarrow 3s^23p^5(\mathrm{^2P^o_{3/2}})3d \; \mathrm{^2[3/2]^o_1}$),
and
carbon lines at $\lambda =$ 
909.48 nm ($2s^22p3s\;^3\mathrm{P^o}_2 \leftarrow 2s^2 2p3p\;^3\mathrm{P}_2$),
1069.125 nm ($2s^2 2p 3s\; ^3\mathrm{P^o}_2 \leftarrow 2s^2 2p 3p\; ^3\mathrm{D}_3$), 
and 1454.25 nm ($2s^2 2p 3s\; ^1\mathrm{P^o}_1 \leftarrow 2s^2 2p 3p\; ^1\mathrm{P_1}$), respectively.

\begin{figure}
    \includegraphics[width=7.5cm]{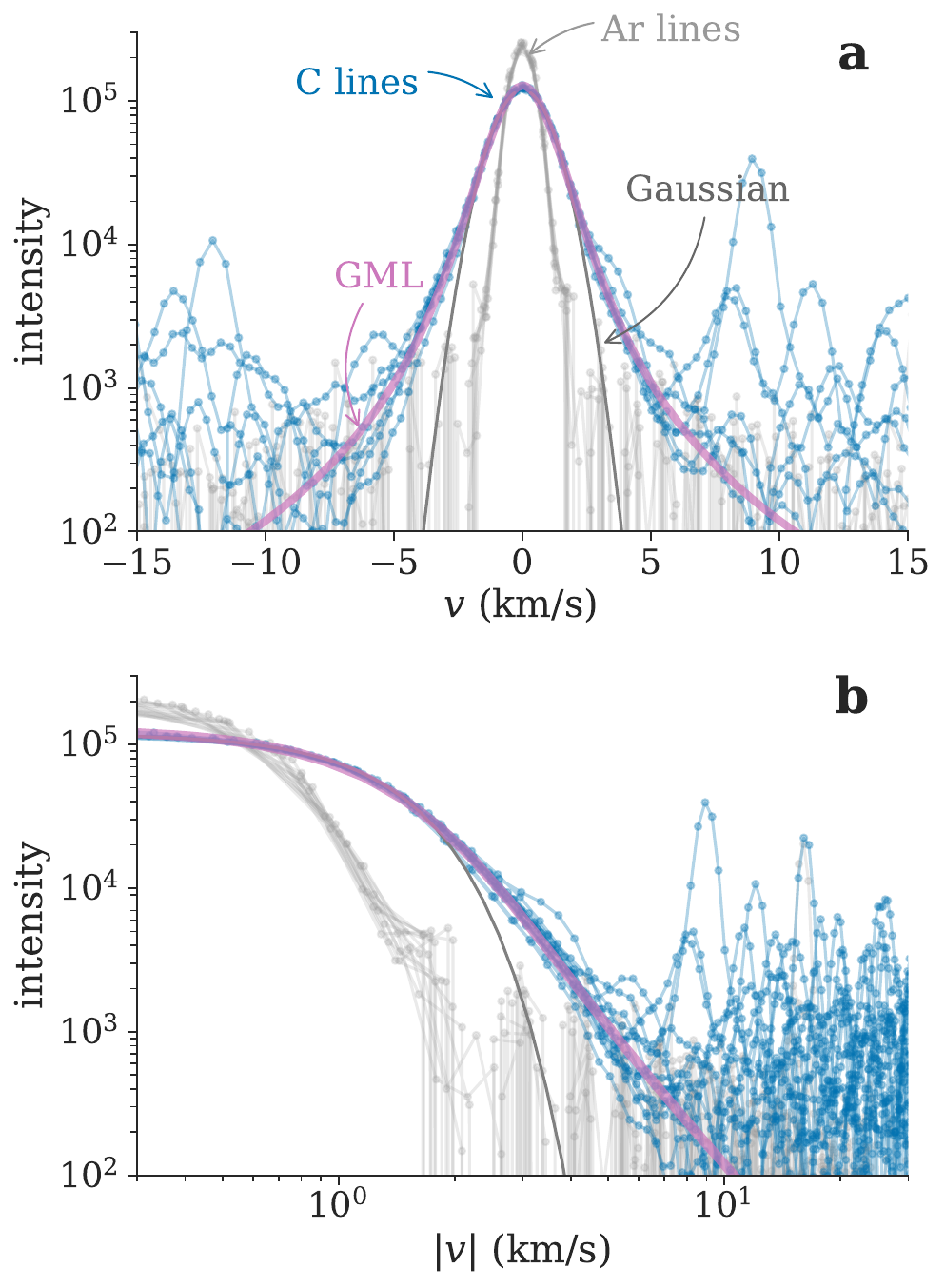}
    \caption{%
    Emission spectra of neutral carbon atoms measured for carbon-nitrogen-argon plasma at the National Solar Observatory~\cite{NSO}.
    $\lambda_0 = $ 
    909.48, 1068.31, 1068.54, 1069.12, 1070.73, 1072.95, and 1454.25 nm lines are plotted with the area normalized to unity.
    The thin and bold curves show the best fit by the Gaussian and the GML velocity distribution, respectively.
    Eight neutral argon emission spectra 
    ($\lambda_0 = $ 1243.93, 1295.67, 1327.26, 1331.32, 1336.71, 1350.42, 1362.27 and 1371.86 nm) 
    are also shown by gray points, which indicate the negligible effect of the instrumental profile on the carbon profiles.
}
    \label{fig:full_nso_C}
\end{figure}

The observed line width of the argon lines is 5.2 pm. 
This is consistent with the convolution of the argon Doppler width at room temperature (3.7 pm) and the instrumental width.
The oscillation seen in the argon line wings, i.e., the side robe, originates from the instrumental function of this spectrometer.
Since this spectrometer is based on Fourier transform principles, the spectrum is affected by the window function, such as the sinc function $\frac{\sin\left(\pi \Delta \lambda / \delta_\lambda \right)}{\pi \Delta \lambda / \delta_\lambda}$ for a rectangular window, where $\delta_\lambda$ is the spectral resolution.
Although other line broadenings, such as Doppler broadening, average out this oscillation in the instrumental side robes, this is still apparent in the argon lines because of their similar Doppler widths to the instrumental width. 
This effect is negligible for the carbon lines owing to their larger Doppler widths.

\Fref{fig:full_nso_C} shows all the emission lines of neutral carbon atoms in the wavelength range 900--1600 nm that are intense enough and isolated from other lines, i.e.,
909.48 nm ($2s^2 2p3s\; ^3\mathrm{P^o_2} \leftarrow 2s^2 2p3p\;\mathrm{^3P_2}$), 
1068.31 nm ($2s^2 2p3s\; \mathrm{^3P^o_1} \leftarrow 2s^2 2p3p\;\mathrm{^3D_2}$), 
1068.54 nm ($2s^22p3s\; \mathrm{^3P^o_0} \leftarrow 2s^2 2p 3p\;\mathrm{^3D_1} $), 
1069.12 nm ($2s^22p3s\; \mathrm{^3P^o_2} \leftarrow 2s^2 2p3p \; \mathrm{^3D_3}$),
1070.73 nm ($2s^22p3s\; \mathrm{^3P^o_1} \leftarrow 2s^2 2p3p\; \mathrm{^3D_1}$), 
1072.95 nm ($2s^22p3s\; \mathrm{^3P^o_2} \leftarrow 2s^2 2p3p\; \mathrm{^3D_2}$), and 
1454.25 nm ($2s^2 2p 3s\; ^1\mathrm{P^o_1} \leftarrow 2s^2 2p 3p\; ^1\mathrm{P_1}$).
They are plotted as functions of the velocity along the sight line, converted by $v = c\Delta \lambda / \lambda_0$, and with their areas normalized.
All the emission lines have the same profile.
Since the Stark broadening and the pressure broadening have different sensitivities depending on the transition, these effects may be negligible. 
From the neutral argon emission lines observed at the same time (gray points in \fref{fig:full_nso_C}), it is found that the instrumental broadening is also negligible on the carbon line profiles.
It is reasonable to assume that the Doppler broadening is dominant that reflects the velocity distribution of ground-state neutral carbon atoms.

The best-fit by a single Gaussian to the carbon lines is shown by a thin curve in \fref{fig:full_nso_C}.
The carbon profile show significant tails compared with a Gaussian function, indicating a strong nonthermality of their velocity distribution.
In a log-log plot (\fref{fig:full_nso_C}~(c)), the velocity distribution is close to a power-law in a large-$|v|$ region.

Carbon atoms in plasmas may also experience heating due to molecular dissociation, entropy production through elastic collision, and energy dissipation by the wall collision.
The GML velocity distribution may be a good candidate to describe this nonthermal velocity distribution.
The bold curve shows the best fit by the GML velocity distribution. 
The entire profile is well represented by this function and from the fit $\alpha = 0.78 \pm 0.01$ is obtained.

\begin{figure}
    \includegraphics[width=7.5cm]{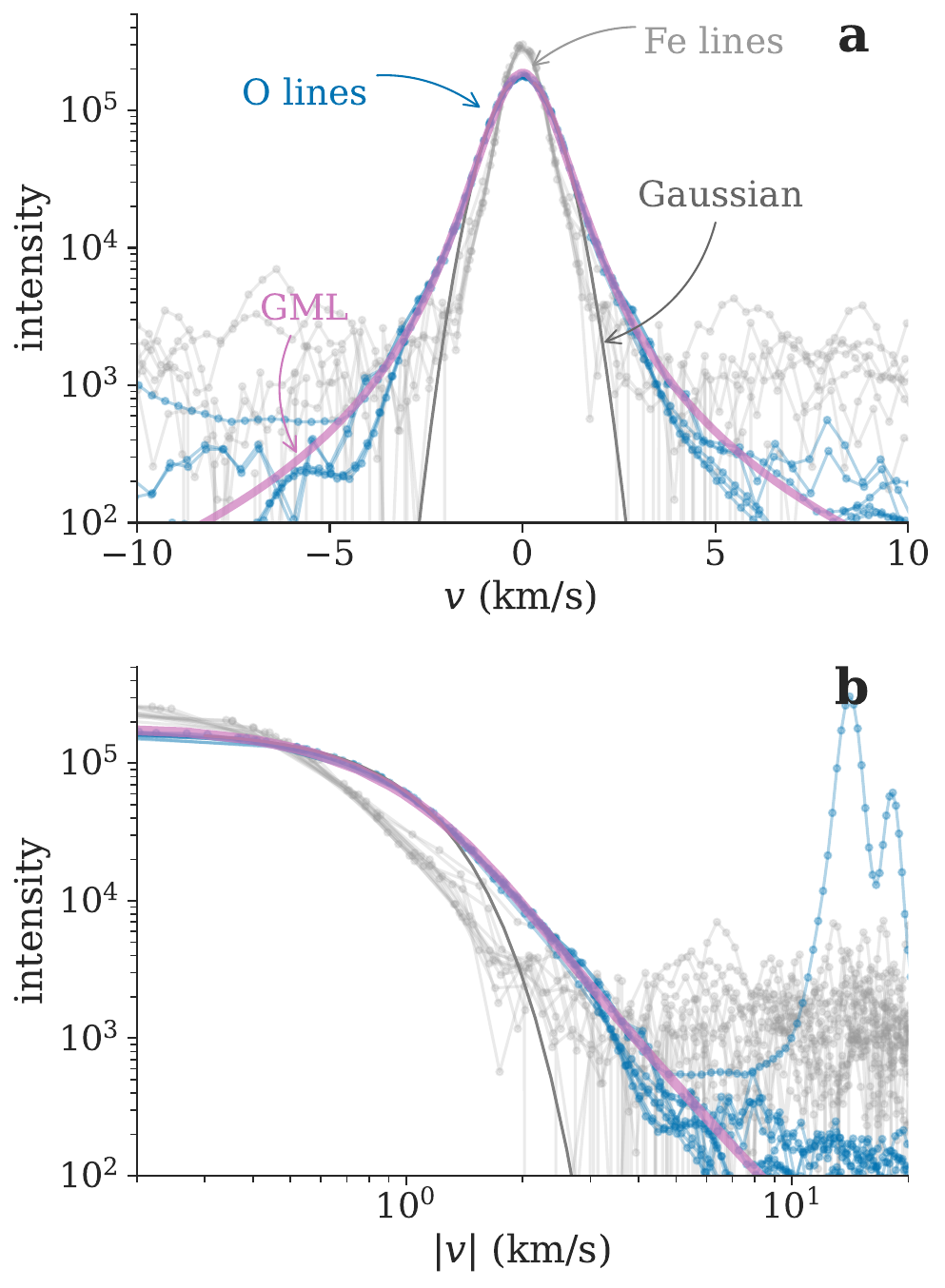}
    \caption{%
    Emission spectra of neutral oxygen atoms measured for iron--helium--carbon-monoxide mixture plasma at the National Solar Observatory~\cite{NSO}.
    $\lambda_0 = $ 
    777.19, 777.42, 777.54, 844.68, 1129.77, and 1130.24 nm lines are plotted with the area normalized to unity.
    The thin and bold curves show the best fit by the Gaussian and the GML velocity distribution, respectively.
    Seven neutral iron emission spectra 
    ($\lambda_0 = $ 744.58, 758.60, 793.71, 794.58,
    851.41, 886.69, and 973.86 nm) 
    are also shown by gray points, which indicate the negligible effect of the instrumental profile on the oxygen profiles.
    }
    \label{fig:full_nso_O}
\end{figure}

Similar spectrum for oxygen atoms can be also found in the same data archive (file \texttt{801015R0.100} in Ref.~\cite{NSO}).
\Fref{fig:full_nso_O} shows the same plot but for an iron--helium--carbon-monoxide mixture plasma. 
Blue points represent the emission line profiles from neutral oxygen,
777.19 nm ($2s^22p^3(\mathrm{^4S^o})3s\; \mathrm{^5S^o_2} \leftarrow 2s^22p^3(\mathrm{^4S^o})3p\; \mathrm{^5P_3}$),  
777.42 nm ($2s^22p^3(\mathrm{^4S^o})3s\; \mathrm{^5S^o_2} \leftarrow 2s^22p^3(\mathrm{^4S^o})3p\; \mathrm{^5P_2}$),  
777.54 nm ($2s^22p^3(\mathrm{^4S^o})3s\; \mathrm{^5S^o_2} \leftarrow 2s^22p^3(\mathrm{^4S^o})3p\; \mathrm{^5P_1}$), 
844.68 nm ($2s^22p^3(\mathrm{^4S^o})3s\; \mathrm{^3S^o_1} \leftarrow 2s^22p^3(\mathrm{^4S^o})3p\; \mathrm{^3P_1}$),
1129.77 nm ($2s^22p^3(\mathrm{^4S^o})3p\; \mathrm{^5P_2} \leftarrow 2s^22p^3(\mathrm{^4S^o})4s\; \mathrm{^5S^o_2}$),
and  
1130.24 nm ($2s^22p^3(\mathrm{^4S^o})3p\; \mathrm{^5P_3} \leftarrow 2s^22p^3(\mathrm{^4S^o})4s\; \mathrm{^5S^o_2}$).
They are plotted as functions of the velocity along the sight line and with their area normalized unity.
All the profiles show a similar distribution.
Also, the comparison with neutral iron lines (
    744.58, 758.60, 793.71, 794.58,
    851.41, 886.69, and 973.86 nm)
shown by gray points in \fref{fig:full_nso_O} indicates the negligible effect by the instrumental broadening.
As similar to the previous discussion, this suggests that the Doppler effect is the dominant broadening mechanism for these lines.

The thin solid curve shows the best-fit Gaussian to these profile. 
The observed neutral oxygen spectra shows significant wing intensity than the Gaussian.
In a log-log plot, it is close to a power-law distribution.
The bold curves in \fref{fig:full_nso_O} show the best fit by the GML velocity distribution. 
The entire profile is well represented by this function, with $\alpha = 0.77 \pm 0.01$ from the fit.

\section{Summary and Discussions}
In this work, an application of the collisional-energy-cascade model to the nonthermal velocity distribution of radical atoms in plasmas has been proposed, which consists of a heat source at the high energy limit, entropy production by elastic collisions, and energy dissipation.
The generalized Mittag-Leffler (GML) distribution is shown to represent a steady-state velocity distribution for this model.

This model is compared with a direct molecular dynamics simulation.
The simulated velocity distribution is well represented by the GML distribution.
Furthermore, the stability parameter in the GML distribution, which indicates the fractional importance of the entropy production against the energy dissipation, is consistent with the theoretical prediction.

The Doppler profiles spectroscopically observed for neutral radical atoms in plasmas are also compared with this model.
They have much stronger wing intensities than a Gaussian function, similar to the molecular dynamics simulation.
The GML velocity distribution well represents the observed velocity distributions, indicating that the collisional energy cascade is universally seen in plasmas.

Historically, nonthermal velocity distributions have been analyzed by two-temperature Maxwellians.
One of the advantages of the GML distribution is that this distribution is based on kinetic theory. The physical knowledge of the system, particularly the stability parameter representing the degree of nonthermality, can be obtained by fitting the GML distribution to the observed spectrum.
Furthermore, as the GML distribution has fewer adjustable parameters (the velocity scale and the stability parameter) than the two-temperature Maxwellian (two velocity scales and the fractions of the two Maxwellians), the GML distribution is more robust.
The GML distribution may provide a new spectroscopy tool to analyze spectra showing nonthermal Doppler profiles. 

Despite strong support by kinetic theory, the GML distribution has some limitations. 
The two-temperature model uses Gaussians, which are favorable when taking instrumental broadening into account, as the convolution of two Gaussians becomes another Gaussian.    
The convolution of the Gaussian-like instrumental function to the GML distribution is not analytically tractable. The use of a look-up table may be a possible extension to consider the instrumental function.

The physics model presented here is not always applicable. 
A nonthermal velocity distribution caused by different mechanisms may not be analyzed by the GML distribution, e.g., when the wall temperature and the heat source temperature are similar, or when emission from the two separate locations with different temperatures are superimposed.
Furthermore, the GML distribution is only applicable $\alpha \gtrsim 0.5$, because of the approximation to derive \eref{eq:mittagleffler}.
If the energy dissipation is dominant compared with elastic collisions, the another treatment would be necessary.
Establishment of the analytical distribution for a strongly dissipative system is left for future work.

\begin{acknowledgments}
    This work was supported by the U.S. D.O.E contract DE-AC05-00OR22725.
    An anonimous person with the username \textit{vitamin d}, who gave us an essential suggestion in \url{https://mathoverflow.net/questions/401835/} is also appreciated.
    Also, the author thanks fruitful comments from Dr. Maeyama (Nagoya University), and Dr. Shiba (University of Tokyo).
\end{acknowledgments}

\bibliography{refs}

\providecommand{\noopsort}[1]{}\providecommand{\singleletter}[1]{#1}%
\begin{thebibliography}{33}%
\makeatletter
\providecommand \@ifxundefined [1]{%
 \@ifx{#1\undefined}
}%
\providecommand \@ifnum [1]{%
 \ifnum #1\expandafter \@firstoftwo
 \else \expandafter \@secondoftwo
 \fi
}%
\providecommand \@ifx [1]{%
 \ifx #1\expandafter \@firstoftwo
 \else \expandafter \@secondoftwo
 \fi
}%
\providecommand \natexlab [1]{#1}%
\providecommand \enquote  [1]{``#1''}%
\providecommand \bibnamefont  [1]{#1}%
\providecommand \bibfnamefont [1]{#1}%
\providecommand \citenamefont [1]{#1}%
\providecommand \href@noop [0]{\@secondoftwo}%
\providecommand \href [0]{\begingroup \@sanitize@url \@href}%
\providecommand \@href[1]{\@@startlink{#1}\@@href}%
\providecommand \@@href[1]{\endgroup#1\@@endlink}%
\providecommand \@sanitize@url [0]{\catcode `\\12\catcode `\$12\catcode
  `\&12\catcode `\#12\catcode `\^12\catcode `\_12\catcode `\%12\relax}%
\providecommand \@@startlink[1]{}%
\providecommand \@@endlink[0]{}%
\providecommand \url  [0]{\begingroup\@sanitize@url \@url }%
\providecommand \@url [1]{\endgroup\@href {#1}{\urlprefix }}%
\providecommand \urlprefix  [0]{URL }%
\providecommand \Eprint [0]{\href }%
\providecommand \doibase [0]{http://dx.doi.org/}%
\providecommand \selectlanguage [0]{\@gobble}%
\providecommand \bibinfo  [0]{\@secondoftwo}%
\providecommand \bibfield  [0]{\@secondoftwo}%
\providecommand \translation [1]{[#1]}%
\providecommand \BibitemOpen [0]{}%
\providecommand \bibitemStop [0]{}%
\providecommand \bibitemNoStop [0]{.\EOS\space}%
\providecommand \EOS [0]{\spacefactor3000\relax}%
\providecommand \BibitemShut  [1]{\csname bibitem#1\endcsname}%
\let\auto@bib@innerbib\@empty
\bibitem [{\citenamefont {Vrhovac}\ \emph {et~al.}(1991)\citenamefont
  {Vrhovac}, \citenamefont {Radovanov}, \citenamefont {Bzeni{\'{c}}},
  \citenamefont {Petrovi{\'{c}}},\ and\ \citenamefont
  {Jelenkovi{\'{c}}}}]{Vrhovac1991}%
  \BibitemOpen
  \bibfield  {author} {\bibinfo {author} {\bibfnamefont {S.}~\bibnamefont
  {Vrhovac}}, \bibinfo {author} {\bibfnamefont {S.}~\bibnamefont {Radovanov}},
  \bibinfo {author} {\bibfnamefont {S.}~\bibnamefont {Bzeni{\'{c}}}}, \bibinfo
  {author} {\bibfnamefont {Z.}~\bibnamefont {Petrovi{\'{c}}}}, \ and\ \bibinfo
  {author} {\bibfnamefont {B.}~\bibnamefont {Jelenkovi{\'{c}}}},\ }\href
  {\doibase 10.1016/0301-0104(91)90021-K} {\bibfield  {journal} {\bibinfo
  {journal} {Chemical Physics}\ }\textbf {\bibinfo {volume} {153}},\ \bibinfo
  {pages} {233} (\bibinfo {year} {1991})}\BibitemShut {NoStop}%
\bibitem [{\citenamefont {Amorim}\ \emph {et~al.}(2000)\citenamefont {Amorim},
  \citenamefont {Baravian},\ and\ \citenamefont {Jolly}}]{Amorim2000}%
  \BibitemOpen
  \bibfield  {author} {\bibinfo {author} {\bibfnamefont {J.}~\bibnamefont
  {Amorim}}, \bibinfo {author} {\bibfnamefont {G.}~\bibnamefont {Baravian}}, \
  and\ \bibinfo {author} {\bibfnamefont {J.}~\bibnamefont {Jolly}},\ }\href
  {\doibase 10.1088/0022-3727/33/9/201} {\bibfield  {journal} {\bibinfo
  {journal} {Journal of Physics D: Applied Physics}\ }\textbf {\bibinfo
  {volume} {33}},\ \bibinfo {pages} {R51} (\bibinfo {year} {2000})}\BibitemShut
  {NoStop}%
\bibitem [{\citenamefont {Samm}\ \emph {et~al.}(1989)\citenamefont {Samm},
  \citenamefont {Bogen}, \citenamefont {Hartwig}, \citenamefont {Hintz},
  \citenamefont {H{\"{o}}thker}, \citenamefont {Lie}, \citenamefont
  {Pospieszczyk}, \citenamefont {Rusb{\"{u}}ldt}, \citenamefont {Schweer},\
  and\ \citenamefont {Yu}}]{Samm1989}%
  \BibitemOpen
  \bibfield  {author} {\bibinfo {author} {\bibfnamefont {U.}~\bibnamefont
  {Samm}}, \bibinfo {author} {\bibfnamefont {P.}~\bibnamefont {Bogen}},
  \bibinfo {author} {\bibfnamefont {H.}~\bibnamefont {Hartwig}}, \bibinfo
  {author} {\bibfnamefont {E.}~\bibnamefont {Hintz}}, \bibinfo {author}
  {\bibfnamefont {K.}~\bibnamefont {H{\"{o}}thker}}, \bibinfo {author}
  {\bibfnamefont {Y.}~\bibnamefont {Lie}}, \bibinfo {author} {\bibfnamefont
  {A.}~\bibnamefont {Pospieszczyk}}, \bibinfo {author} {\bibfnamefont
  {D.}~\bibnamefont {Rusb{\"{u}}ldt}}, \bibinfo {author} {\bibfnamefont
  {B.}~\bibnamefont {Schweer}}, \ and\ \bibinfo {author} {\bibfnamefont
  {Y.}~\bibnamefont {Yu}},\ }\href {\doibase 10.1016/0022-3115(89)90255-9}
  {\bibfield  {journal} {\bibinfo  {journal} {Journal of Nuclear Materials}\
  }\textbf {\bibinfo {volume} {162-164}},\ \bibinfo {pages} {24} (\bibinfo
  {year} {1989})}\BibitemShut {NoStop}%
\bibitem [{\citenamefont {Hey}\ \emph {et~al.}(1999)\citenamefont {Hey},
  \citenamefont {Chu},\ and\ \citenamefont {Hintz}}]{Hey1999}%
  \BibitemOpen
  \bibfield  {author} {\bibinfo {author} {\bibfnamefont {J.~D.}\ \bibnamefont
  {Hey}}, \bibinfo {author} {\bibfnamefont {C.~C.}\ \bibnamefont {Chu}}, \ and\
  \bibinfo {author} {\bibfnamefont {E.}~\bibnamefont {Hintz}},\ }\href
  {\doibase 10.1088/0953-4075/32/14/321} {\bibfield  {journal} {\bibinfo
  {journal} {Journal of Physics B: Atomic, Molecular and Optical Physics}\
  }\textbf {\bibinfo {volume} {32}},\ \bibinfo {pages} {3555} (\bibinfo {year}
  {1999})}\BibitemShut {NoStop}%
\bibitem [{\citenamefont {Shikama}\ \emph {et~al.}(2004)\citenamefont
  {Shikama}, \citenamefont {Kado}, \citenamefont {Zushi}, \citenamefont
  {Iwamae},\ and\ \citenamefont {Tanaka}}]{Shikama2004}%
  \BibitemOpen
  \bibfield  {author} {\bibinfo {author} {\bibfnamefont {T.}~\bibnamefont
  {Shikama}}, \bibinfo {author} {\bibfnamefont {S.}~\bibnamefont {Kado}},
  \bibinfo {author} {\bibfnamefont {H.}~\bibnamefont {Zushi}}, \bibinfo
  {author} {\bibfnamefont {A.}~\bibnamefont {Iwamae}}, \ and\ \bibinfo {author}
  {\bibfnamefont {S.}~\bibnamefont {Tanaka}},\ }\href {\doibase
  10.1063/1.1783877} {\bibfield  {journal} {\bibinfo  {journal} {Physics of
  Plasmas}\ }\textbf {\bibinfo {volume} {11}},\ \bibinfo {pages} {4701}
  (\bibinfo {year} {2004})}\BibitemShut {NoStop}%
\bibitem [{\citenamefont {Sasaki}\ \emph {et~al.}(2009)\citenamefont {Sasaki},
  \citenamefont {Okumura},\ and\ \citenamefont {Asaoka}}]{Sasaki2009-et}%
  \BibitemOpen
  \bibfield  {author} {\bibinfo {author} {\bibfnamefont {K.}~\bibnamefont
  {Sasaki}}, \bibinfo {author} {\bibfnamefont {Y.}~\bibnamefont {Okumura}}, \
  and\ \bibinfo {author} {\bibfnamefont {R.}~\bibnamefont {Asaoka}},\ }\href
  {\doibase 10.1155/2010/627571} {\bibfield  {journal} {\bibinfo  {journal}
  {International Journal of Spectroscopy}\ }\textbf {\bibinfo {volume} {2010}}
  (\bibinfo {year} {2009}),\ 10.1155/2010/627571}\BibitemShut {NoStop}%
\bibitem [{\citenamefont {Corrigan}(1965)}]{Corrigan1965}%
  \BibitemOpen
  \bibfield  {author} {\bibinfo {author} {\bibfnamefont {S.~J.~B.}\
  \bibnamefont {Corrigan}},\ }\href {\doibase 10.1063/1.1696701} {\bibfield
  {journal} {\bibinfo  {journal} {The Journal of Chemical Physics}\ }\textbf
  {\bibinfo {volume} {43}},\ \bibinfo {pages} {4381} (\bibinfo {year}
  {1965})}\BibitemShut {NoStop}%
\bibitem [{\citenamefont {Hey}\ \emph {et~al.}(2004)\citenamefont {Hey},
  \citenamefont {Chu}, \citenamefont {Mertens}, \citenamefont {Brezinsek},\
  and\ \citenamefont {Unterberg}}]{Hey2004}%
  \BibitemOpen
  \bibfield  {author} {\bibinfo {author} {\bibfnamefont {J.~D.}\ \bibnamefont
  {Hey}}, \bibinfo {author} {\bibfnamefont {C.~C.}\ \bibnamefont {Chu}},
  \bibinfo {author} {\bibfnamefont {P.}~\bibnamefont {Mertens}}, \bibinfo
  {author} {\bibfnamefont {S.}~\bibnamefont {Brezinsek}}, \ and\ \bibinfo
  {author} {\bibfnamefont {B.}~\bibnamefont {Unterberg}},\ }\href {\doibase
  10.1088/0953-4075/37/12/010} {\bibfield  {journal} {\bibinfo  {journal}
  {Journal of Physics B: Atomic, Molecular and Optical Physics}\ }\textbf
  {\bibinfo {volume} {37}},\ \bibinfo {pages} {2543} (\bibinfo {year}
  {2004})}\BibitemShut {NoStop}%
\bibitem [{\citenamefont {Scarlett}\ \emph {et~al.}(2017)\citenamefont
  {Scarlett}, \citenamefont {Tapley}, \citenamefont {Fursa}, \citenamefont
  {Zammit}, \citenamefont {Savage},\ and\ \citenamefont {Bray}}]{Scarlett2017}%
  \BibitemOpen
  \bibfield  {author} {\bibinfo {author} {\bibfnamefont {L.~H.}\ \bibnamefont
  {Scarlett}}, \bibinfo {author} {\bibfnamefont {J.~K.}\ \bibnamefont
  {Tapley}}, \bibinfo {author} {\bibfnamefont {D.~V.}\ \bibnamefont {Fursa}},
  \bibinfo {author} {\bibfnamefont {M.~C.}\ \bibnamefont {Zammit}}, \bibinfo
  {author} {\bibfnamefont {J.~S.}\ \bibnamefont {Savage}}, \ and\ \bibinfo
  {author} {\bibfnamefont {I.}~\bibnamefont {Bray}},\ }\href {\doibase
  10.1103/PhysRevA.96.062708} {\bibfield  {journal} {\bibinfo  {journal}
  {PHYSICAL REVIEW A}\ }\textbf {\bibinfo {volume} {96}},\ \bibinfo {pages}
  {62708} (\bibinfo {year} {2017})}\BibitemShut {NoStop}%
\bibitem [{\citenamefont {McConkey}\ \emph {et~al.}(2008)\citenamefont
  {McConkey}, \citenamefont {Malone}, \citenamefont {Johnson}, \citenamefont
  {Winstead}, \citenamefont {McKoy},\ and\ \citenamefont
  {Kanik}}]{McConkey2008}%
  \BibitemOpen
  \bibfield  {author} {\bibinfo {author} {\bibfnamefont {J.}~\bibnamefont
  {McConkey}}, \bibinfo {author} {\bibfnamefont {C.}~\bibnamefont {Malone}},
  \bibinfo {author} {\bibfnamefont {P.}~\bibnamefont {Johnson}}, \bibinfo
  {author} {\bibfnamefont {C.}~\bibnamefont {Winstead}}, \bibinfo {author}
  {\bibfnamefont {V.}~\bibnamefont {McKoy}}, \ and\ \bibinfo {author}
  {\bibfnamefont {I.}~\bibnamefont {Kanik}},\ }\href {\doibase
  10.1016/j.physrep.2008.05.001} {\bibfield  {journal} {\bibinfo  {journal}
  {Physics Reports}\ }\textbf {\bibinfo {volume} {466}},\ \bibinfo {pages} {1}
  (\bibinfo {year} {2008})}\BibitemShut {NoStop}%
\bibitem [{\citenamefont {Starikovskiy}(2015)}]{Starikovskiy2015}%
  \BibitemOpen
  \bibfield  {author} {\bibinfo {author} {\bibfnamefont {A.~Y.}\ \bibnamefont
  {Starikovskiy}},\ }\href {\doibase 10.1098/rsta.2014.0343} {\bibfield
  {journal} {\bibinfo  {journal} {Philosophical Transactions of the Royal
  Society A: Mathematical, Physical and Engineering Sciences}\ }\textbf
  {\bibinfo {volume} {373}},\ \bibinfo {pages} {20140343} (\bibinfo {year}
  {2015})}\BibitemShut {NoStop}%
\bibitem [{\citenamefont {Sommerer}\ and\ \citenamefont
  {Kushner}(1991)}]{Sommerer1991}%
  \BibitemOpen
  \bibfield  {author} {\bibinfo {author} {\bibfnamefont {T.~J.}\ \bibnamefont
  {Sommerer}}\ and\ \bibinfo {author} {\bibfnamefont {M.~J.}\ \bibnamefont
  {Kushner}},\ }\href {\doibase 10.1063/1.349579} {\bibfield  {journal}
  {\bibinfo  {journal} {Journal of Applied Physics}\ }\textbf {\bibinfo
  {volume} {70}},\ \bibinfo {pages} {1240} (\bibinfo {year}
  {1991})}\BibitemShut {NoStop}%
\bibitem [{\citenamefont {Ponomarev}\ and\ \citenamefont
  {Aleksandrov}(2017)}]{Ponomarev2017}%
  \BibitemOpen
  \bibfield  {author} {\bibinfo {author} {\bibfnamefont {A.~A.}\ \bibnamefont
  {Ponomarev}}\ and\ \bibinfo {author} {\bibfnamefont {N.~L.}\ \bibnamefont
  {Aleksandrov}},\ }\href {\doibase 10.1088/1361-6595/aa5f41} {\bibfield
  {journal} {\bibinfo  {journal} {Plasma Sources Science and Technology}\
  }\textbf {\bibinfo {volume} {26}},\ \bibinfo {pages} {044003} (\bibinfo
  {year} {2017})}\BibitemShut {NoStop}%
\bibitem [{\citenamefont {Ben-Naim}\ and\ \citenamefont
  {Machta}(2005)}]{Ben-Naim2005-rd}%
  \BibitemOpen
  \bibfield  {author} {\bibinfo {author} {\bibfnamefont {E.}~\bibnamefont
  {Ben-Naim}}\ and\ \bibinfo {author} {\bibfnamefont {J.}~\bibnamefont
  {Machta}},\ }\href {\doibase 10.1103/PhysRevLett.94.138001} {\bibfield
  {journal} {\bibinfo  {journal} {Physical review letters}\ }\textbf {\bibinfo
  {volume} {94}},\ \bibinfo {pages} {138001} (\bibinfo {year}
  {2005})}\BibitemShut {NoStop}%
\bibitem [{\citenamefont {Ben-Naim}\ \emph {et~al.}(2005)\citenamefont
  {Ben-Naim}, \citenamefont {Machta},\ and\ \citenamefont
  {Machta}}]{Ben-Naim2005-uz}%
  \BibitemOpen
  \bibfield  {author} {\bibinfo {author} {\bibfnamefont {E.}~\bibnamefont
  {Ben-Naim}}, \bibinfo {author} {\bibfnamefont {B.}~\bibnamefont {Machta}}, \
  and\ \bibinfo {author} {\bibfnamefont {J.}~\bibnamefont {Machta}},\ }\href
  {\doibase 10.1103/PhysRevE.72.021302} {\bibfield  {journal} {\bibinfo
  {journal} {Physical Review E}\ }\textbf {\bibinfo {volume} {72}},\ \bibinfo
  {pages} {021302} (\bibinfo {year} {2005})}\BibitemShut {NoStop}%
\bibitem [{\citenamefont {Kang}\ \emph {et~al.}(2010)\citenamefont {Kang},
  \citenamefont {Machta},\ and\ \citenamefont {Ben-Naim}}]{Kang2010-tk}%
  \BibitemOpen
  \bibfield  {author} {\bibinfo {author} {\bibfnamefont {W.}~\bibnamefont
  {Kang}}, \bibinfo {author} {\bibfnamefont {J.}~\bibnamefont {Machta}}, \ and\
  \bibinfo {author} {\bibfnamefont {E.}~\bibnamefont {Ben-Naim}},\ }\href
  {\doibase 10.1209/0295-5075/91/34002} {\bibfield  {journal} {\bibinfo
  {journal} {EPL}\ }\textbf {\bibinfo {volume} {91}},\ \bibinfo {pages} {34002}
  (\bibinfo {year} {2010})}\BibitemShut {NoStop}%
\bibitem [{\citenamefont {Fujii}(2022)}]{Fujii2022}%
  \BibitemOpen
  \bibfield  {author} {\bibinfo {author} {\bibfnamefont {K.}~\bibnamefont
  {Fujii}},\ }\href {\doibase 10.48550/arXiv.2210.06193} {\bibfield  {journal}
  {\bibinfo  {journal} {Submitted to Physical Review Letters}\ } (\bibinfo
  {year} {2022}),\ 10.48550/arXiv.2210.06193}\BibitemShut {NoStop}%
\bibitem [{\citenamefont {Futcher}\ and\ \citenamefont
  {Hoare}(1980)}]{Futcher1980-ey}%
  \BibitemOpen
  \bibfield  {author} {\bibinfo {author} {\bibfnamefont {E.~J.}\ \bibnamefont
  {Futcher}}\ and\ \bibinfo {author} {\bibfnamefont {M.~R.}\ \bibnamefont
  {Hoare}},\ }\href {\doibase 10.1016/0375-9601(80)90042-0} {\bibfield
  {journal} {\bibinfo  {journal} {Physics letters. A}\ }\textbf {\bibinfo
  {volume} {75}},\ \bibinfo {pages} {443} (\bibinfo {year} {1980})}\BibitemShut
  {NoStop}%
\bibitem [{\citenamefont {Hendriks}\ and\ \citenamefont
  {Ernst}(1982)}]{Hendriks1982-cw}%
  \BibitemOpen
  \bibfield  {author} {\bibinfo {author} {\bibfnamefont {E.~M.}\ \bibnamefont
  {Hendriks}}\ and\ \bibinfo {author} {\bibfnamefont {M.~H.}\ \bibnamefont
  {Ernst}},\ }\href@noop {} {\bibfield  {journal} {\bibinfo  {journal} {Physica
  A: Statistical Mechanics and its Applications}\ }\textbf {\bibinfo {volume}
  {112}},\ \bibinfo {pages} {119} (\bibinfo {year} {1982})}\BibitemShut
  {NoStop}%
\bibitem [{\citenamefont {Futcher}\ and\ \citenamefont
  {Hoare}(1983)}]{Futcher1983-sf}%
  \BibitemOpen
  \bibfield  {author} {\bibinfo {author} {\bibfnamefont {E.~J.}\ \bibnamefont
  {Futcher}}\ and\ \bibinfo {author} {\bibfnamefont {M.~R.}\ \bibnamefont
  {Hoare}},\ }\href {\doibase 10.1016/0378-4371(83)90047-X} {\bibfield
  {journal} {\bibinfo  {journal} {Physica A: Statistical Mechanics and its
  Applications}\ }\textbf {\bibinfo {volume} {122}},\ \bibinfo {pages} {516}
  (\bibinfo {year} {1983})}\BibitemShut {NoStop}%
\bibitem [{\citenamefont {Haubold}\ \emph {et~al.}(2011)\citenamefont
  {Haubold}, \citenamefont {Mathai},\ and\ \citenamefont
  {Saxena}}]{haubold_mittag-leffler_2011}%
  \BibitemOpen
  \bibfield  {author} {\bibinfo {author} {\bibfnamefont {H.~J.}\ \bibnamefont
  {Haubold}}, \bibinfo {author} {\bibfnamefont {A.~M.}\ \bibnamefont {Mathai}},
  \ and\ \bibinfo {author} {\bibfnamefont {R.~K.}\ \bibnamefont {Saxena}},\
  }\href {\doibase 10.1155/2011/298628} {\bibfield  {journal} {\bibinfo
  {journal} {Journal of Applied Mathematics}\ ,\ \bibinfo {pages} {Art. ID
  298628, 51}} (\bibinfo {year} {2011})}\BibitemShut {NoStop}%
\bibitem [{\citenamefont {Barabesi}\ \emph {et~al.}(2016)\citenamefont
  {Barabesi}, \citenamefont {Cerasa}, \citenamefont {Cerioli},\ and\
  \citenamefont {Perrotta}}]{barabesi_new_2016}%
  \BibitemOpen
  \bibfield  {author} {\bibinfo {author} {\bibfnamefont {L.}~\bibnamefont
  {Barabesi}}, \bibinfo {author} {\bibfnamefont {A.}~\bibnamefont {Cerasa}},
  \bibinfo {author} {\bibfnamefont {A.}~\bibnamefont {Cerioli}}, \ and\
  \bibinfo {author} {\bibfnamefont {D.}~\bibnamefont {Perrotta}},\ }\href
  {\doibase 10.1214/16-EJS1214} {\bibfield  {journal} {\bibinfo  {journal}
  {Electronic Journal of Statistics}\ }\textbf {\bibinfo {volume} {10}},\
  \bibinfo {pages} {3871} (\bibinfo {year} {2016})},\ \bibinfo {note}
  {publisher: Institute of Mathematical Statistics and Bernoulli
  Society}\BibitemShut {NoStop}%
\bibitem [{\citenamefont {Korolev}\ \emph {et~al.}(2020)\citenamefont
  {Korolev}, \citenamefont {Gorshenin},\ and\ \citenamefont
  {Zeifman}}]{korolev_mixture_2020}%
  \BibitemOpen
  \bibfield  {author} {\bibinfo {author} {\bibfnamefont {V.}~\bibnamefont
  {Korolev}}, \bibinfo {author} {\bibfnamefont {A.}~\bibnamefont {Gorshenin}},
  \ and\ \bibinfo {author} {\bibfnamefont {A.}~\bibnamefont {Zeifman}},\ }\href
  {\doibase 10.1007/s10958-020-04755-8} {\bibfield  {journal} {\bibinfo
  {journal} {Journal of Mathematical Sciences}\ }\textbf {\bibinfo {volume}
  {246}},\ \bibinfo {pages} {503} (\bibinfo {year} {2020})}\BibitemShut
  {NoStop}%
\bibitem [{\citenamefont {Massey}(1934)}]{Massey1934}%
  \BibitemOpen
  \bibfield  {author} {\bibinfo {author} {\bibfnamefont {H.~S.~W.}\
  \bibnamefont {Massey}},\ }\href {\doibase 10.1098/rspa.1934.0042} {\bibfield
  {journal} {\bibinfo  {journal} {Proceedings of the Royal Society of London.
  Series A, Containing Papers of a Mathematical and Physical Character}\
  }\textbf {\bibinfo {volume} {144}},\ \bibinfo {pages} {188} (\bibinfo {year}
  {1934})}\BibitemShut {NoStop}%
\bibitem [{\citenamefont {Flannery}(2006)}]{Flannery2006}%
  \BibitemOpen
  \bibfield  {author} {\bibinfo {author} {\bibfnamefont {M.}~\bibnamefont
  {Flannery}},\ }in\ \href {\doibase 10.1007/978-0-387-26308-3_45} {\emph
  {\bibinfo {booktitle} {Springer Handbook of Atomic, Molecular, and Optical
  Physics}}}\ (\bibinfo  {publisher} {Springer New York},\ \bibinfo {address}
  {New York, NY},\ \bibinfo {year} {2006})\ pp.\ \bibinfo {pages}
  {659--691}\BibitemShut {NoStop}%
\bibitem [{\citenamefont {{Ito}}\ \emph {et~al.}(1985)\citenamefont {{Ito}},
  \citenamefont {{Tabata}}, \citenamefont {{Itoh}}, \citenamefont {{Morita}},
  \citenamefont {{Kato}},\ and\ \citenamefont {{Tawara}}}]{Ito1985-px}%
  \BibitemOpen
  \bibfield  {author} {\bibinfo {author} {\bibnamefont {{Ito}}}, \bibinfo
  {author} {\bibnamefont {{Tabata}}}, \bibinfo {author} {\bibnamefont
  {{Itoh}}}, \bibinfo {author} {\bibnamefont {{Morita}}}, \bibinfo {author}
  {\bibnamefont {{Kato}}}, \ and\ \bibinfo {author} {\bibnamefont {{Tawara}}},\
  }\href@noop {} {\bibfield  {journal} {\bibinfo  {journal} {IPPJ-AM41,
  Institute of Plasma}\ } (\bibinfo {year} {1985})}\BibitemShut {NoStop}%
\bibitem [{\citenamefont {Thompson}\ \emph {et~al.}(2022)\citenamefont
  {Thompson}, \citenamefont {Aktulga}, \citenamefont {Berger}, \citenamefont
  {Bolintineanu}, \citenamefont {Brown}, \citenamefont {Crozier}, \citenamefont
  {in~'t Veld}, \citenamefont {Kohlmeyer}, \citenamefont {Moore}, \citenamefont
  {Nguyen}, \citenamefont {Shan}, \citenamefont {Stevens}, \citenamefont
  {Tranchida}, \citenamefont {Trott},\ and\ \citenamefont {Plimpton}}]{LAMMPS}%
  \BibitemOpen
  \bibfield  {author} {\bibinfo {author} {\bibfnamefont {A.~P.}\ \bibnamefont
  {Thompson}}, \bibinfo {author} {\bibfnamefont {H.~M.}\ \bibnamefont
  {Aktulga}}, \bibinfo {author} {\bibfnamefont {R.}~\bibnamefont {Berger}},
  \bibinfo {author} {\bibfnamefont {D.~S.}\ \bibnamefont {Bolintineanu}},
  \bibinfo {author} {\bibfnamefont {W.~M.}\ \bibnamefont {Brown}}, \bibinfo
  {author} {\bibfnamefont {P.~S.}\ \bibnamefont {Crozier}}, \bibinfo {author}
  {\bibfnamefont {P.~J.}\ \bibnamefont {in~'t Veld}}, \bibinfo {author}
  {\bibfnamefont {A.}~\bibnamefont {Kohlmeyer}}, \bibinfo {author}
  {\bibfnamefont {S.~G.}\ \bibnamefont {Moore}}, \bibinfo {author}
  {\bibfnamefont {T.~D.}\ \bibnamefont {Nguyen}}, \bibinfo {author}
  {\bibfnamefont {R.}~\bibnamefont {Shan}}, \bibinfo {author} {\bibfnamefont
  {M.~J.}\ \bibnamefont {Stevens}}, \bibinfo {author} {\bibfnamefont
  {J.}~\bibnamefont {Tranchida}}, \bibinfo {author} {\bibfnamefont
  {C.}~\bibnamefont {Trott}}, \ and\ \bibinfo {author} {\bibfnamefont {S.~J.}\
  \bibnamefont {Plimpton}},\ }\href {\doibase 10.1016/j.cpc.2021.108171}
  {\bibfield  {journal} {\bibinfo  {journal} {Comp. Phys. Comm.}\ }\textbf
  {\bibinfo {volume} {271}},\ \bibinfo {pages} {108171} (\bibinfo {year}
  {2022})}\BibitemShut {NoStop}%
\bibitem [{\citenamefont {Baroody}(1961)}]{Baroody1961-vx}%
  \BibitemOpen
  \bibfield  {author} {\bibinfo {author} {\bibfnamefont {E.~M.}\ \bibnamefont
  {Baroody}},\ }\href@noop {} {\bibfield  {journal} {\bibinfo  {journal} {The
  Physics of Fluids}\ }\textbf {\bibinfo {volume} {4}},\ \bibinfo {pages}
  {1182} (\bibinfo {year} {1961})}\BibitemShut {NoStop}%
\bibitem [{\citenamefont {Amorim}\ \emph
  {et~al.}(1996{\natexlab{a}})\citenamefont {Amorim}, \citenamefont
  {Baravian},\ and\ \citenamefont {Sultan}}]{Amorim1996}%
  \BibitemOpen
  \bibfield  {author} {\bibinfo {author} {\bibfnamefont {J.}~\bibnamefont
  {Amorim}}, \bibinfo {author} {\bibfnamefont {G.}~\bibnamefont {Baravian}}, \
  and\ \bibinfo {author} {\bibfnamefont {G.}~\bibnamefont {Sultan}},\ }\href
  {\doibase 10.1063/1.116293} {\bibfield  {journal} {\bibinfo  {journal}
  {Applied Physics Letters}\ }\textbf {\bibinfo {volume} {68}},\ \bibinfo
  {pages} {1915} (\bibinfo {year} {1996}{\natexlab{a}})}\BibitemShut {NoStop}%
\bibitem [{NSO()}]{NSO}%
  \BibitemOpen
  \href@noop {} {\enquote {\bibinfo {title} {{NSO's historical archive}},}\
  }\bibinfo {howpublished} {\url{https://nso.edu/data/historical-archive/}},\
  \bibinfo {note} {[Online; accessed 01-Feb-2021]}\BibitemShut {NoStop}%
\bibitem [{\citenamefont {Amorim}\ \emph
  {et~al.}(1996{\natexlab{b}})\citenamefont {Amorim}, \citenamefont
  {Baravian},\ and\ \citenamefont {Sultan}}]{Amorim1996-gj}%
  \BibitemOpen
  \bibfield  {author} {\bibinfo {author} {\bibfnamefont {J.}~\bibnamefont
  {Amorim}}, \bibinfo {author} {\bibfnamefont {G.}~\bibnamefont {Baravian}}, \
  and\ \bibinfo {author} {\bibfnamefont {G.}~\bibnamefont {Sultan}},\ }\href
  {\doibase 10.1063/1.116293} {\bibfield  {journal} {\bibinfo  {journal}
  {Applied physics letters}\ }\textbf {\bibinfo {volume} {68}},\ \bibinfo
  {pages} {1915} (\bibinfo {year} {1996}{\natexlab{b}})}\BibitemShut {NoStop}%
\bibitem [{int(1999)}]{international1999iaea}%
  \BibitemOpen
  \href
  {https://www.iaea.org/publications/4710/atomic-and-plasma-material-interaction-data-for-fusion}
  {\emph {\bibinfo {title} {Atomic and Plasma^^e2^^80^^93Material Interaction
  Data for Fusion}}},\ \bibinfo {series} {Atomic and Plasma^^e2^^80^^93Material
  Interaction Data for Fusion}\ No.~\bibinfo {number} {8}\ (\bibinfo
  {publisher} {INTERNATIONAL ATOMIC ENERGY AGENCY},\ \bibinfo {address}
  {Vienna},\ \bibinfo {year} {1999})\BibitemShut {NoStop}%
\bibitem [{\citenamefont {Mankodi}\ \emph {et~al.}(2020)\citenamefont
  {Mankodi}, \citenamefont {Bhandarkar},\ and\ \citenamefont
  {Myong}}]{Mankodi2020-ve}%
  \BibitemOpen
  \bibfield  {author} {\bibinfo {author} {\bibfnamefont {T.~K.}\ \bibnamefont
  {Mankodi}}, \bibinfo {author} {\bibfnamefont {U.~V.}\ \bibnamefont
  {Bhandarkar}}, \ and\ \bibinfo {author} {\bibfnamefont {R.~S.~e.}\
  \bibnamefont {Myong}},\ }\href {\doibase 10.1063/1.5141148} {\bibfield
  {journal} {\bibinfo  {journal} {Physics of fluids}\ }\textbf {\bibinfo
  {volume} {32}},\ \bibinfo {pages} {036102} (\bibinfo {year}
  {2020})}\BibitemShut {NoStop}%
\end{thebibliography}%

    
\end{document}